# Ultrafast light–electron modulation for free-electron quantum optics: fundamentals and applications

Zirui Zhao[1]†, Ruoyu Chen[1,2]†, Changying Li[1], Yongcheng Ding[3], Zhaopin Chen[4,5,6], Xinbing Song[7], Shuaishuai Sun[8], Ye Tian[2]*, Kangpeng Wang[9]*, Yiming Pan[1]*

[1] *State Key Laboratory of Quantum Functional Materials, School of Physical Science and Technology and Center for Transformative Science, ShanghaiTech University, Shanghai 200031, China*
[2] *State Key Laboratory of Ultra-intense Laser Science and Technology, Shanghai Institute of Optics and Fine Mechanics, Chinese Academy of Sciences, Shanghai, China*
[3] *Department of Physical Chemistry, University of the Basque Country UPV/EHU, Apartado 644, 48080 Bilbao, Spain*
[4] *Department of Physics, Technion – Israel Institute of Technology, Haifa 3200003, Israel*
[5] *Solid State Institute, Technion – Israel Institute of Technology, Haifa 3200003, Israel*
[6] *The Helen Diller Quantum Center, Technion – Israel Institute of Technology, Haifa3200003, Israel*
[7] *School of Physics, Beijing Institute of Technology, Beijing 100081, China*
[8] *Beijing National Laboratory for Condensed Matter Physics, Institute of Physics, Chinese Academy of Sciences, Beijing 100190, China*
[9] *Shanghai Institute of Optics and Fine Mechanics, Chinese Academy of Sciences, Shanghai 201800, China.*

**Abstract**

Coherent optical modulation of free-electron wavepackets offers a route to ultrafast beam steering, energy exchange, temporal compression, and quantum-state shaping. Recent advances in photon-induced near-field electron microscopy (PINEM), ultrafast Kapitza-Dirac effect (KDE) and dielectric laser acceleration (DLA) have demonstrated strong light–electron coupling, but longitudinal energy modulation, transverse momentum transfer, and wavepacket-dependent diffraction are still commonly described with separate models. Here, we develop a unified Light–Electron Modulation (LEM) framework that treats optical near fields as programmable phase modulators for free-electron wavepackets. Within this framework, we identify longitudinal and transverse modulation channels—analogous to electro-optic phase modulation and acousto-optic diffraction of light, respectively—and show how their operation depends on wavepacket extent, recoil, phase matching, and optical-field geometry. This formulation organizes PINEM, Raman–Nath and Bragg diffraction, DLA-like acceleration, and anomalous PINEM as limiting regimes of a common modulation picture. The resulting framework provides design principles for free-electron modulators that can control energy spectra, temporal density, transverse wavefronts, and electron–photon quantum interfaces, establishing LEM as a device-level platform for programmable femtosecond-to-attosecond matter-wave control.

## Introduction

Controlled manipulation of free electrons underlies modern particle optics, electron microscopy, and accelerator science. In conventional electron-optical systems, the electron is often treated as a classical point charge whose trajectory is shaped by externally applied electromagnetic fields. This particle-based description has enabled mature beamline components for steering, focusing, filtering, and acceleration, but it does not fully capture the coherent wavepacket dynamics that emerge when free electrons interact with optical fields.

In this classical framework, electron motion is described by the Lorentz force, $m\ddot{r} = -e(\boldsymbol{E} + \boldsymbol{v} \times \boldsymbol{B})$. Static and quasi-static $\boldsymbol{E}$ and $\boldsymbol{B}$ fields provide a versatile toolkit for controlling electron position, momentum, energy, and arrival time through magnetic lenses, Wien filters[1,2], deflectors, and related electron-optical elements[3]. These components remain indispensable for beam transport and phase-space control. However, operating in static or low-frequency regimes, these conventional components cannot exploit the optical-frequency phase modulation of a coherent electron wavepacket. This limitation motivates the light–electron modulation (LEM) framework developed in this work.

A quantum description instead treats the electron as a matter wave that can diffract, interfere, and acquire field-dependent phase[4]. The interaction with external electromagnetic fields is introduced through minimal coupling. In the nonrelativistic limit, the time-dependent Schrödinger equation can be written as $i\hbar \frac{\partial}{\partial t}\psi = \left[\frac{1}{2m}(\boldsymbol{p} + e\boldsymbol{A})^2 - e\phi\right]\psi$, where the vector potential $\boldsymbol{A}$ and scalar potential $\phi$ couple the external field to the electron wavefunction. In the LEM picture, the electron source prepares an incident quantum electron wavepacket (QEW), and the optical interaction region imprints a controllable phase or momentum modulation on this wavepacket (Fig. 1).

The same electromagnetic potentials also describe optical fields, with $\boldsymbol{E} = -\partial\boldsymbol{A}/\partial t - \nabla\phi$ $and$ $\boldsymbol{B} = \nabla \times \boldsymbol{A}$. This common formal structure connects static, radio-frequency, and optical electron modulation, but the operating constraints change substantially at optical frequencies. Optical near fields can provide subcycle timing and nanoscale spatial gradients, while also imposing requirements on phase matching, field confinement, synchronization, and quantum recoil. LEM therefore treats the optical field as a programmable high-frequency input that modulates the phase, energy, or momentum of a free-electron wavepacket, extending coherent electron beam control into the femtosecond and attosecond regimes.

The recent development of ultrafast lasers and nanophotonic structures makes this optical-frequency control experimentally accessible. Femtosecond optical pulses provide phase-stable timing references, while structured near fields can localize the interaction to nanometer-scale regions and enhance coupling to a passing electron. These capabilities allow light to act not only as a probe of electron or material dynamics, but also as an active drive that reshapes the electron wavepacket.

This control strategy is conceptually related to state engineering in atomic, molecular, and optical physics, where lasers are used to prepare, couple, and read out quantum states. For free electrons, however, the relevant degrees of freedom are continuous—spanning energy, time, and momentum, and the interaction is strongly influenced by wavepacket length, recoil, and propagation-induced dispersion. These differences mean that LEM requires a dedicated regime map rather than a direct transfer of bound-state intuition. Section 2 develops such a map by classifying diffraction and modulation regimes according to wavepacket character and phase-matching conditions.

A particle-like limit of optical electron control is realized in dielectric laser acceleration (DLA)[5–7]. In DLA, phase-synchronous optical near fields in nanofabricated dielectric structures transfer energy to electrons over subwavelength interaction regions, enabling large accelerating gradients. For the purposes of LEM, however, the central role of DLA is not only acceleration, but also its demonstration that an optical near field can impose a phase-dependent energy shift on a localized electron wavepacket.

In our framework, DLA corresponds to the localized-wavepacket limit of optical modulation. When the electron wavepacket is short compared with the optical period, it samples a well-defined phase of the driving field and acquires a continuous energy shift or velocity chirp rather than a resolved multiphoton sideband comb. This limit connects classical acceleration, phase-dependent energy modulation, and temporal bunching within a single LEM description (Figs. 2 and 3).

The complementary wave-like limit is exemplified by Photon-Induced Near-Field Electron Microscopy (PINEM) [8–10]. In PINEM, an extended electron wavepacket interacts coherently with an optical near field and develops discrete energy sidebands separated by the photon energy. This sideband structure shows that the near field acts as a phase-modulating interaction rather than only as a stroboscopic probe [11]. Within LEM, PINEM is therefore identified as an extended-wavepacket Raman–Nath regime of optical phase modulation (Figs. 2 and 3).

In a coherently driven PINEM interaction, the outgoing electron spectrum contains sidebands spaced by $\hbar\omega$, and their populations are described by Bessel-function weights[10,11]. This response can be interpreted as a quantum walk on a synthetic energy ladder, where optical modulation couples neighboring electron energy states. In the LEM architecture of Fig. 1, the same process is viewed at the device level as an input QEW, an optically driven near-field modulator, and an energy-resolved output. This establishes the swift-electron Raman–Nath limit used in the regime map of Fig. 2.

These two limits are connected by the quantum state of the driving light and by the longitudinal structure of the electron wavepacket. Different photon statistics can imprint distinct sideband statistics on the electron spectrum, enabling electron-based probes of quantum light[12–14]. Independently, the electron wavepacket length controls whether the interaction appears as a DLA-like continuous energy modulation or as a PINEM-like discrete sideband comb[15–17]. Pre-interaction free-space dispersion adds a further control parameter by chirping the QEW and producing anomalous PINEM (APINEM) spectra[18]. These dependencies motivate a unified model in which photon statistics, wavepacket extent, recoil, and dispersion are treated as parameters of the same modulation process.

This modulation viewpoint also applies to transverse and time-domain control schemes, including Kapitza-Dirac diffraction[19–22], attosecond streaking[23,24], and radio-frequency beam shaping[25,26]. While these schemes differ in frequency, geometry, and readout observable, they all redistribute electron energy or momentum through structured electromagnetic fields. We therefore classify LEM devices by operational geometry and coupling channel: longitudinal versus transverse modulation, linear versus ponderomotive interaction, and traveling-wave versus standing-wave configurations. This classification, developed in Section 3, converts a set of apparently distinct light–electron effects into design regimes for programmable free-electron modulation.

At the quantum level, these modulation processes connect to free-electron quantum optics (FEQO), where free electrons exchange energy, momentum, and quantum information with photonic or material excitations[27]. In this context, LEM provides a controllable interface for mapping selected optical properties, including phase, amplitude, photon statistics, and mode structure, onto measurable electron degrees of freedom. This interface perspective motivates the quantum and reciprocal-modulation applications discussed in later sections.

The LEM framework is formulated by comparing the information-bearing degrees of freedom of a structured optical field with those of a free-electron wavefunction. Specifically, a classical optical mode is written as $E(\boldsymbol{r},t) = Re\{\chi_P A e^{-i\omega t} e^{ik_z z} e^{im\theta}\}$, while the electron wavefunction can be expressed as:

$$\psi(\boldsymbol{r},t) = \chi_S \varphi_0 e^{i\phi(r,t)} = \chi_S \varphi_0 e^{-iEt/\hbar} e^{ip_z z/\hbar} e^{im\theta}.$$

This comparison highlights the degrees of freedom addressable in light–electron modulation. Optical fields encode information across polarization ($\chi_P$), amplitude ($A$), frequency ($\omega$), phase, and orbital angular momentum ($m$) , whereas free-electron wavefunctions are modulated through amplitude ($\varphi_0$), energy ($E$), phase ($\phi$), and spin ($\chi_S$) , and momentum-space structure. The analogy is useful but not exact: the accessibility of these channels depends on interaction geometry, wavepacket length, recoil, and propagation-induced chirp. Therefore, a central requirement of the LEM model is to treat the transition from localized, particle-like acceleration to extended-wavepacket quantum diffraction within the same formalism.

At the device level, LEM operates as a modulation-and-readout architecture. A structured optical field supplies the modulation waveform, an incident QEW serves as the matter-wave carrier, and the output energy spectrum, temporal density, transverse momentum distribution, or phase-space profile provides the readout. The objective of this work is to derive a unified modelling framework that connects these measurable outputs to the underlying light–electron interaction Hamiltonian and classifies the resulting operating regimes.

The remainder of the article is organized as follows. Section 2 derives the LEM phase-modulation model from the minimal-coupling Hamiltonian and classifies diffraction regimes according to wavepacket extent, recoil, and phase matching. Section 3 develops an operational device taxonomy based on longitudinal versus transverse modulation, linear versus ponderomotive coupling, and traveling- versus standing-wave geometries. Section 4 examines material and nanophotonic platforms for near-field confinement and phase-matched coupling. Section 5 analyzes representative LEM functions, including energy shifting, temporal bunching, beam shaping, and quantum-state sorting. Section 6 discusses engineering constraints such as coherence preservation, space charge, dissipation, and strong-coupling requirements, before Section 7 summarizes the implications for integrated free-electron modulation devices.

This work focuses on the modelling principles and device regimes of light–electron modulation rather than on a comprehensive review of free-electron quantum optics. We use selected theoretical and experimental results to define limiting cases, validate the regime classification, and identify design constraints for LEM devices. Related mechanisms, including free-electron bound-electron resonant interaction[28–31], are therefore discussed only where they clarify the scope of the LEM framework.

## Theory and diffraction regimes

### A. Phase-modulation model of light–electron interaction

We first formulate LEM as a device-level input–interaction–readout model. An electron source prepares an incident quantum electron wavepacket (QEW), an optically driven interaction region imprints a phase modulation, and a downstream spectrometer projects the outgoing state onto measurable energy, time, or momentum observables. In the semiclassical limit, the optical near field acts as a phase mask on the electron wavefunction: $\psi_{out}(z,t) = e^{i\phi(z,t)}\psi_{in}(z,t)$.

As the first analytically tractable limit, we consider a longitudinally extended QEW that approaches a plane-wave state and is subject to sinusoidal optical phase modulation. For an incident monochromatic component with kinetic energy $E_0$, we have $\psi_{\text{in}}(t) = e^{-iE_0 t/\hbar}$. Using the co-moving time coordinate $t - z/v_0 \rightarrow t$, the outgoing state is[11]:

$$\psi_{out}(t) = e^{i2|g|\sin(\omega_L t)}e^{-\frac{iE_0 t}{\hbar}} = \sum_{n=-\infty}^{\infty} J_n(2|g|)e^{-\frac{i(E_0 - n\hbar\omega_L)t}{\hbar}}, \tag{1}$$

Here, $2|g|$ denotes the optical modulation strength and $\omega_L$ is the angular frequency of the driving field. Projecting the outgoing wavefunction onto the energy basis gives:

$$P(E) = |\mathcal{F}[\psi_{out}(t)]|^2 = \sum_{n=-\infty}^{\infty} |J_n(2|g|)|^2 \delta(E - E_0 + n\hbar\omega_L) = \sum_{n=-\infty}^{\infty} P_n \,. \tag{2}$$

Thus, the sideband population is $P_n = |J_n(2|g|)|^2$. The integer n labels photon sidebands with energies $E_n = E_0 - n\hbar\omega_L$, where $n > 0$ corresponds to electron energy loss and $n < 0$ to energy gain. Sinusoidal phase modulation populates a discrete synthetic energy ladder with spacing $\hbar\omega_L$ and Bessel-function weights. This energy-ladder picture provides the Raman–Nath reference limit used throughout the regime map. In organizing the diffraction landscape, we distinguish between swift electrons—typical of PINEM (e.g., $E_0 = 200\,keV, \beta = v/c = 0.7$)—and slow electrons (e.g., $E_0 = 100\,eV, \beta = 0.02$), where quantum recoil and second-order dispersion become prominent (Fig. 2).

This phase-modulation form is directly analogous to conventional electro-optic modulation (EOM)[32]. In an electro-optic modulator, a radio-frequency drive imprints a time-dependent phase on an optical carrier, producing $Ae^{-i\omega t + i\beta \sin(\Omega t)}$ and optical sidebands. In LEM, the optical near field plays the role of the modulation drive, while the free-electron wavefunction acts as the matter-wave carrier. The analogy is useful because both systems convert temporal phase modulation into discrete spectral components, although their carriers, coupling mechanisms, and energy scales differ. To connect the phase-modulation model to microscopic coupling, we introduce $H_I = \alpha p \cdot A + \beta A^2$. The $p \cdot A$ term defines the linear coupling channel, whereas the $A^2$ term governs the ponderomotive coupling channel.

For a nonrelativistic electron in vacuum, the coefficients of the linear and ponderomotive coupling terms follow directly from minimal coupling. Taking the electron charge as $q = -e$, the minimal-coupling Hamiltonian expands as:

$$H = \frac{(p + eA)^2}{2m} = \frac{p^2}{2m} + \frac{e}{m} p \cdot A + \frac{e^2}{2m} A^2, \tag{3}$$

This Hamiltonian separates the interaction into a linear $p \cdot A$ channel and a quadratic $A^2$ channel. The linear term usually dominates when the electron velocity has a component along the optical vector potential and the field amplitude is moderate. The ponderomotive term becomes important for strong optical fields or for geometries in which $p \cdot A$ is suppressed, such as orthogonal coupling configurations. In this regime, the $A^2$ interaction imprints a phase grating on the electron wavefunction, leading to Kapitza–Dirac diffraction, high-order diffraction channels, and Rabi-like population oscillations [33].

The two coupling channels also provide a useful analogy with electro-optic modulation. The linear $p \cdot A$ term acts analogously to a first-order electro-optic response, whereas the quadratic $A^2$ term corresponds to a Kerr-like response, as the ponderomotive phase scales with field intensity. This analogy reflects an operational correspondence rather than a material nonlinearity of the free electron: p·A and A² provide linear-field and intensity-dependent modulation channels, respectively, for the electron wavepacket. Acousto-optic modulation provides a second analogy. In an acousto-optic modulator (AOM), a moving refractive-index grating $n(x,t) = n_0 + n_1 \cos(Gx - \Omega t)$ produces frequency shifts and angular selectivity. In LEM, an engineered optical near field creates an analogous spatiotemporal phase grating for free electrons. When phase-matching conditions are satisfied, the electron exchanges energy and momentum with this moving grating.

The AOM analogy motivates the distinction between the Raman–Nath and Bragg diffraction regimes. In the Raman–Nath limit, the interaction acts as a thin phase grating and populates many diffraction or energy orders. In the Bragg limit, an extended, phase-matched interaction selects a small number of resonant orders and supports coherent population oscillations. For free electrons, these diffraction limits correspond to the swift and slow velocity regimes, respectively. We therefore use the electron velocity as one axis of the LEM regime map in Fig. 2 to represent this transition. The other axis is the longitudinal character of the incident QEW. Combining these axes allows PINEM, DLA-like acceleration, APINEM, spectral breathing, spectral shearing, and Bragg diffraction to be treated as limiting regimes of the same light–electron modulation model.

**Quantum electron wavepacket characterization.** To describe the crossover between particle-like acceleration and wave-like diffraction, we introduce the quantum electron wavepacket (QEW) description, where the intrinsic longitudinal wavepacket width determines whether the electron samples a local optical phase or interferes across multiple optical cycles. We define the incident QEW at the modulator entrance by a Gaussian momentum-space envelope:

$$\psi_0(p) = \left(2\pi\sigma_p^2\right)^{-1/4} \exp\left[-\frac{(p-p_0)^2}{4\sigma_p^2}\right] \tag{4}$$

where $p_0$ is the central momentum and $\sigma_p$ is the momentum spread. The free-space evolution of this wavepacket is determined by the relativistic dispersion relation expanded around $p_0$ to second order: $E_p = c\sqrt{m_0^2c^2 + p^2} = E_0 + v_0(p-p_0) + \frac{(p-p_0)^2}{2m^*}$, where $E_0$ is the central kinetic energy, $v_0 = \beta c$ is the group velocity, and $m^* = \gamma^3 m_0$ is the relativistic longitudinal effective mass. Constructing the real-space wavefunction via the Fourier transform yields the dispersive QEW profile:

$$\psi_0(z,t) = \frac{e^{i(p_0 z - E_0 t)/\hbar}}{(2\pi\sigma_z^2)^{1/4}\sqrt{1+i\xi t}} \exp\left[-\frac{(z-v_0 t)^2}{4\sigma_z^2(1+i\xi t)}\right] \tag{5}$$

where $\sigma_z = \hbar/(2\sigma_p)$ defines the intrinsic longitudinal width, and $\xi = \hbar/(2m^*\sigma_z^2)$ characterizes the vacuum-induced chirp rate governing the free-space expansion of the wavepacket. To ground this formalism in typical ultrafast transmission electron microscopy (UTEM) parameters, we consider an electron pulse with a central kinetic energy $E_0 = 200$ keV ( $\gamma \approx 1.4$, $\beta \approx 0.7$) and an intrinsic energy spread $\sigma_E = 1$ eV yields an intrinsic longitudinal width of $\sigma_{z0} \approx 69$ nm and a transform-limited pulse duration of $\sigma_t = \sigma_z/v_0 = \hbar/(2\sigma_E) \approx 0.33$ fs. Crucially, this intrinsic quantum duration is nearly two orders of magnitude shorter than the macroscopic electron pulse durations ($\sim 30$ fs) conventionally measured at the sample level[34].

Experimentally measured ultrafast electron pulses exhibit temporal durations and energy spreads that vastly exceed these intrinsic quantum limits, owing to both the coherent vacuum dispersion (chirp) of the single-electron wavefunction and the incoherent statistical fluctuations of the many-particle ensemble. Formally, the ensemble-averaged macroscopic pulse duration $\langle \Delta t \rangle$ and energy spread $\langle \Delta E \rangle$ are decomposed as:

$$\begin{aligned} \langle \Delta_\mathrm{t} \rangle &= \sqrt{\sigma_t^2(1+\xi^2 t_0^2) + \Delta_{t,en}^2} \\ \langle \Delta_E \rangle &= \sqrt{\sigma_E^2 + \Delta_{E,en}^2} \end{aligned} \tag{6}$$

where $t_0$ denotes the pre-interaction drift time, and $\Delta_{t,\mathrm{en}}$ and $\Delta_{E,\mathrm{en}}$ represent the statistical ensemble spreads. These incoherent contributions originate from initial photoemission kinematics, space-charge (Coulomb) repulsion, and classical aberrations induced during transverse focusing and filtering by the magnetic condenser lens [35–37]. Consequently, the macroscopic phase-space volume (typically $\langle \Delta t \rangle \langle \Delta E \rangle \approx 30$ eV · fs) exceeds the fundamental Heisenberg uncertainty limit ( $\sigma_t \sigma_E = \hbar/2 \approx 0.33$ eV · fs) by nearly two orders of magnitude, obscuring the underlying transform-limited quantum state.

The relevant wavepacket parameter for coherent LEM is therefore the intrinsic longitudinal size $\sigma_z$, rather than the macroscopic pulse duration alone. Comparing $\sigma_z$ with the optical wavelength λ defines two limiting responses. In the localized-wavepacket limit, $\sigma_z \ll \lambda$, the electron samples a local optical phase and can acquire a continuous DLA-like energy shift or velocity chirp. In the extended-wavepacket limit, $\sigma_z \gg \lambda$, the electron interacts coherently over many optical phases and develops discrete multiphoton sidebands characteristic of PINEM. The following sections derive these limits and extend the classification to chirped and recoil-sensitive regimes.

### B. Swift-electron Raman-Nath/ PINEM Pattern

**PINEM Pattern.** We first derive the standard PINEM response as the swift-electron Raman–Nath limit of the LEM model. In the Coulomb gauge, $\nabla \cdot A = 0$, and with electron charge $q = -e$, the linear interaction term is $H_I = \frac{e}{m} p \cdot A$. For a swift electron crossing a thin optical near-field region, recoil and free-space dispersion during the interaction can be neglected. The vector potential can then be evaluated along the unperturbed trajectory and approximated by $A(r,t) \approx A_{eff} e^{i\omega_L t}$.

The phase shift imprinted on the electron wavefunction is determined by the temporal integration of the interaction Hamiltonian along the unperturbed trajectory. For an incident electron with central momentum $p_0 = mv_0$, the accumulated optical phase $\Delta\phi$ simplifies to an integration over the effective transit time. Utilizing the relationship between the driving electric field and the vector potential, $E = -\partial A/\partial t = E_{opt}\sin(\omega_L t)$, the field amplitude relates as $A_L = E_{opt}/\omega_L$. The resulting accumulated phase is therefore:

$$\Delta\phi(t) = -\frac{eLE_{opt}}{\hbar\omega_L}\cos(\omega_L t) \equiv -2|g|\cos(\omega_L t) \tag{7}$$

where $2|g|$ denotes the dimensionless modulation strength [10,11]. Applying this sinusoidal phase mask, the output wavefunction in the time domain is expanded via the Jacobi–Anger identity:

$$\psi_{out}(t) = e^{-i2|g|cos(\omega_L t)}\psi_{in}(t) = \sum_{n=-\infty}^{+\infty}(-i)^n J_n(2|g|)e^{in\omega_L t}\psi_{in}(t) \tag{8}$$

This expression mathematically encapsulates the multi-photon embedding process, confirming that the interaction operates as a pure sinusoidal phase modulation. Performing a Fourier transform yields the corresponding spectral distribution in the energy domain:

$$\psi_{out}(E) = \sum_{n=-\infty}^{+\infty}(-i)^n J_n(2|g|)\,\delta(E - E_0 + n\hbar\omega_L) \tag{9}$$

This dictates that in the plane-wave Raman–Nath limit, the electron spectrum evolves into a discrete comb of energy sidebands, whose populations are fully determined by Bessel-function statistics, ($P_n = |J_n(2|g|)|^2$).

For a finite incident QEW, the delta-function sidebands broaden into Gaussian momentum components in the co-moving frame. Introducing intrinsic momentum spread $\sigma_p$, the modulated wavefunction becomes:

$$\psi_p = \left(\frac{1}{2\pi\sigma_p^2}\right)^{1/4}\sum_n(-i)^n J_n(2|g|)e^{-in\phi_0}exp\left(-\frac{(p - p_0 + n\delta p)^2}{4\sigma_p^2}\right) \tag{10}$$

with $\delta p = \hbar\omega_L/v_0$ is the momentum spacing associated with one photon-energy exchange, and $p_0 = \gamma m_0 v_0$ is the center electron momentum , and $\phi_0$ is the optical phase at the modulator entrance. Near $p_0$, $E - E_0 \approx v_0(p - p_0)$, so that the longitudinal momentum coordinate can equivalently be represented as an energy coordinate, with $\sigma_E = v_0\sigma_p$ and $\delta E_{ph} = v_0\delta p = \hbar\omega_L$. The superposition of these Gaussian sidebands defines a synthetic momentum–energy lattice (multiphoton embedding) that enables coherent quantum state control and energy-to-density conversion via free-space drift (Fig. 7). Furthermore, as the incident wavepacket narrows in real space toward the point-particle limit, the interference among adjacent sidebands drives the physical transition from the discrete Raman–Nath regime into the continuous DLA and anomalous PINEM (APINEM) limits, confirming that these limiting cases fundamentally manifest as limiting regimes of a unified Raman–Nath diffraction picture.

**DLA-like modulation as the localized-wavepacket limit.** In this regime, $\sigma_p \gg \delta p$, equivalently $\sigma_E \gg \hbar\omega_L$. Adjacent photon-order components overlap strongly, and the discrete PINEM comb approaches a continuous spectral translation. This limit can be evaluated by applying $e^{\varepsilon \partial_E} f(E) = f(E+\varepsilon)$. With the sign convention that $n > 0$ denotes electron energy loss, the energy-domain wavefunction can be written as:

$$\psi_{\text{out}}(E) \approx \sum_{n=-\infty}^{\infty} J_n(2|g|) e^{-in\phi_0} e^{n\hbar\omega_L \partial_E} \psi_{in}(E) \approx e^{2|g|\hbar\omega_L \cos\phi_0 \partial_E} \psi_{in}(E)$$
$$= \psi_{\text{in}}(E + 2|g|\hbar\omega_L \cos\phi_0)$$

Thus, the localized-wavepacket limit yields a deterministic energy transfer:

$$\delta E(\phi_0) = \langle E \rangle_{out} - \mathrm{E}_0 = -2|g|\hbar\omega_L \cos\phi_0 \quad (11)$$

Using the modulation-strength definition $2|g| = \frac{eLE_{opt}}{\hbar\omega_L}$, the phase-dependent energy transfer becomes $\delta E(\phi_0) = -eE_{opt}Lcos\phi_0$. However, DLA-like modulation is not limited to net acceleration. Because the localized QEW samples a singular optical phase $\phi_0$, tuning this phase dictates the operational mode (Fig. 5). At the accelerating ($\phi_0 = \pi$) and decelerating ($\phi_0 = 0$) phases, the spectrum rigidly shifts toward higher or lower energy, respectively[15,38]. In quadrature ($\phi_0 = \pi/2, 3\pi/2$), the net energy shift vanishes, but the optical-field gradient imposes a velocity chirp that drives longitudinal focusing or defocusing. While these dynamics mirror the classical motion of a particle governed by the Lorentz force, this derivation demonstrates that such trajectory manipulation is rooted in quantum multiphoton interference.

Alternatively, this macroscopic energy shift can be obtained from the classical Lorentz-force work along the electron trajectory. Assuming a near-resonant transit where the unperturbed coordinate evolves as $z(t) \approx v_0 t$, the net energy shift imparted by the alternating optical field $E_z(z,t) = E_{opt}\cos(q_z z - \omega_L t - \phi_0)$ is obtained by integrating the force over the effective interaction length $L$:

$$\Delta E = -e\upsilon_0 \int_0^{L/v_0} E_{opt} \cos\big((q_z \upsilon_0 - \omega_L)t - \phi_0\big)\, dt = -eE_{opt}Lsinc\left(\frac{\theta}{2}\right)\cos\left(\phi_0 + \frac{\theta}{2}\right) \quad (12)$$

with $\theta = (\omega_L/v_0 - q_z)L$ parameterizing the accumulated phase mismatch. However, a rigorous quantum-mechanical treatment reveals that this macroscopic acceleration is fundamentally an interference phenomenon. When the electron's intrinsic energy spread vastly exceeds the photon energy quantum ($\sigma_E \gg \hbar\omega_L$), the discrete PINEM-type multiphoton sidebands strongly overlap. This extensive interference collapses the discrete quantum walk into a continuous spectral shift, manifesting macroscopically as a net energy gain or loss. For a Gaussian wavepacket, this continuous momentum transfer is modulated by a quantum exponential decay factor[15]:

$$\Delta E = -2|g|\hbar\omega_L cos\phi_0 e^{-\Gamma_0^2}, \tag{13}$$

where $\Gamma_0 = \hbar\omega_L/2\sigma_E$ acts as an "annihilation factor" governing the visibility of the macroscopic acceleration. In the strict point-particle limit ( $\Gamma_0 \ll 1$ ), this exponential term approaches unity, perfectly recovering the classical net acceleration, shifting the entire Gaussian wavepacket as a rigid body in momentum (energy) space rather than partitioning it into discrete sidebands.

This result gives a clear physical interpretation of the coupling parameter $2|g|$. In the extended-wavepacket limit, $2|g|$ sets the bandwidth of the PINEM comb through Bessel-function weights, whereas in the localized-wavepacket limit, it determines the magnitude of the DLA-like energy translation[39,40]. Thus, sideband generation and classical acceleration are two limiting manifestations of the same phase-modulation strength[17]. The continuous translation limit also suggests an analogy with frequency shearing in electro-optic modulators (EOMs) for pulses shorter than the RF period. By analogy with Spectral Phase Interferometry for Direct Electric-field Reconstruction (SPIDER)[41,42], operating an LEM in this shearing limit provides an interferometric route to reconstructing the spectral amplitude and phase of free-electron wavepackets[43].

**APINEM as highly chirped Raman–Nath limit.** The preceding limits assume that the QEW reaches the interaction region without substantial pre-interaction GVD-induced chirp. Unlike optical pulses in purely phase-modulated architectures, massive free electrons invariably accumulate group-velocity dispersion (GVD) during free-space propagation. A wavepacket that is initially localized near the emission source can therefore expand and become strongly pre-chirped before entering the LEM. The interaction between this pre-chirped QEW and the optical near field defines the anomalous PINEM (APINEM) regime.

In the highly chirped limit, the multiphoton interference no longer produces a rigid DLA-like translation or a standard PINEM comb; instead, the interaction yields pronounced periodic spectral bunching and focusing along the momentum axis. Crucially, the momentum spacing of these APINEM microbunches deviates entirely from the standard photon recoil ($\delta p = \hbar\omega_L/v_0$); rather, it is dispersion-controlled and scales as[18]

$$\delta p_{\mathrm{APINEM}} \propto \frac{\lambda_L}{L_0}. \tag{14}$$

where $\lambda_L$ is the optical wavelength and $L_0$ is the pre-drift distance. Under THz-field modulation, this dispersion-controlled spectral periodicity can synthesize sub-meV spectral features, establishing a robust mechanism for generating highly monochromatic electron energy beams.

If the transit time through the LEM is short compared with the dispersion time scale, dispersion accumulated inside the interaction region can be neglected. Free-space propagation before the LEM acts as state preparation that generates a chirped QEW, while the LEM itself acts as an instantaneous sinusoidal phase mask. Under this separation, the PINEM scattering amplitudes remain $J_n(2g)$, but they are applied to a chirped Gaussian basis:

$$\psi_{\mathrm{p}}(t_0) = \left(\frac{1}{2\pi\sigma_p^2}\right)^{1/4} \sum_n J_n\,(2|g|)e^{-in\phi_0} exp\left(-\frac{(p - p_0 + n\delta p)^2}{4\sigma_p^2(1 + i\xi t_0)^{-1}}\right), \tag{15}$$

where $\xi = 2\sigma_{\mathrm{p}}^2/\mathrm{m}^*\hbar$ is the vacuum chirp coefficient and $\mathrm{t}_0 = L_0/v_0$ is the pre-interaction drift time. If intra-device dispersion becomes non-negligible, this instantaneous phase-mask approximation breaks down, requiring full numerical integration of the time-dependent Schrödinger equation[40].

### C. Slow-electron Bragg diffraction regimes (Few-photon embedding)

**Two-level electron and Rabi oscillation.** For slow electrons, recoil and phase mismatch become comparable to the multiphoton coupling strength, and the instantaneous Raman–Nath phase-mask approximation is no longer sufficient. As the incident momentum decreases, different photon-order components acquire different velocity- and phase-matching conditions, suppressing unrestricted growth of the sideband ladder. The interaction enters the Bragg regime, where only a small set of resonant states remains coupled. In the extreme Bragg limit, the dynamics truncate to two resonant sidebands, forming an effective two-level pseudospin subspace that supports coherent Rabi oscillations [44–47]. If the wavepacket is strongly localized, these broadened sidebands coherently overlap, yielding optical Stern–Gerlach-type diffraction[48–51].

To formalize Bragg-regime dynamics, we expand the modulated wavefunction via the Floquet–Bloch theorem: $\psi(z,t) = \sum_n c_n(t)e^{i(k_n z - \omega_n t)}$, where the sideband energies and momenta scale as $\hbar\omega_n = E_0 + n\hbar\omega_L$ and $\hbar k_n = p_0 + n\hbar q_z$. Substituting this ansatz under the small-recoil approximation ( $q_z \ll p_0/\hbar$ ) maps the quantum dynamics onto tight-binding coupled-mode equations [45,47]:

$$i\hbar\frac{\mathrm{dc_n}}{dt} = \epsilon_n c_n + \kappa c_{n+1} + \kappa^* c_{n-1}, \tag{16}$$

with the energy detuning $\epsilon_n = n\hbar(\omega_L - v_0 q_z) - \frac{n^2\hbar^2 q_z^2}{2\gamma^3 m_0}$. Here, the linear mismatch term represents the phase/velocity mismatch and the quadratic term represents the recoil-induced kinetic dispersion. This dispersion term governs the topological evolution of the synthetic momentum lattice: in the swift-electron limit, it vanishes to recover the Raman–Nath random walk, whereas linear mismatch drives synthetic Bloch oscillations[52,53], and the quadratic scaling imposes a harmonic boundary that naturally confines the multiphoton cascade.

In the deep Bragg regime, the quadratic recoil detuning can isolate an effective two-level subspace. When the incident electron satisfies the Bragg resonance condition, the input state and the first diffracted order become energetically degenerate. Labelling them by pseudo-spin indices $n = \pm 1/2$, their diagonal detunes are $\epsilon/4$, while the nearest off-resonant states $n = \pm 3/2$ lie at $9\epsilon/4$. When the dispersion gap vastly exceeds the coupling strength ( $\epsilon \gg |\kappa|$ ), higher sidebands are adiabatically eliminated, collapsing the infinite coupled equations into a closed two-level system:

$$\begin{cases} i\hbar \dfrac{dc_{1/2}}{dt} = \dfrac{\epsilon}{4} c_{\frac{1}{2}} + \kappa^* c_{-\frac{1}{2}} \\ i\hbar \dfrac{dc_{-1/2}}{dt} = \dfrac{\epsilon}{4} c_{-\frac{1}{2}} + \kappa c_{\frac{1}{2}} \end{cases} \tag{17}$$

Assuming the electron is initially injected into the $|-1/2\rangle$ state, this population transfer yields coherent Rabi oscillations:

$$\begin{cases} P_{\frac{1}{2}}(t) = \sin^2\left(\dfrac{|\kappa| t}{\hbar}\right) \\ P_{-\frac{1}{2}}(t) = \cos^2\left(\dfrac{|\kappa| t}{\hbar}\right) \end{cases} \tag{18}$$

This deterministic, reversible exchange between the discrete momentum states constitutes a light-induced "two-level" electron Rabi oscillation.

Crucially, this isolated $SU(2)$ subspace provides the physical foundation for a genuine free-electron qubit[54–56]. Unlike dynamically rigid Raman–Nath parity qubits, the Bragg-regime qubit allows arbitrary coherent quantum rotations via standard optical pulses (Fig. 6). This deterministic control firmly establishes the free electron as a programmable carrier for quantum information processing.

Relaxing the two-level resonance condition reveals an equally fundamental three-level Bragg regime centered on $n = 0$. For an incident electron matched to the central momentum state, the quadratic dispersion dictates the energy levels: $\epsilon_0 = 0$, $\epsilon_{\pm 1} = \epsilon$, and $\epsilon_{\pm 2} = 4\epsilon$. When the optical coupling is weak $(|\kappa| \ll 4\epsilon)$, the energy barrier at $n = \pm 2$ blocks higher-order scattering. The infinite synthetic lattice thus truncates to the three-level subspace $n \in \{-1, 0, +1\}$, governed by:

$$\begin{cases} i\hbar \dfrac{\mathrm{d}c_0}{dt} = \kappa c_1 + \kappa^* c_{-1} \\ i\hbar \dfrac{\mathrm{d}c_{\pm 1}}{dt} = \epsilon c_{\pm 1} + \kappa^{(*)} c_0 2 \end{cases} \tag{19}$$

This electron wavefunction oscillates between the carrier state $|0\rangle$ and the symmetric sideband superposition $(|-1\rangle + |1\rangle)/\sqrt{2}$. Under THz modulation, this Bragg embedding appears as deterministic spectral breathing (Fig. 2). This regime provides a physical bridge between the unbounded swift-electron random walk and the isolated two-level pseudospin dynamics.

**Optical Stern-Gerlach limit and anomalous Bragg regimes**. Beyond Rabi oscillations, the slow-electron Bragg regime supports an optical Stern–Gerlach (OSG) limit[48]. Under strict two-level truncation, the resonant momentum states form a coherent pseudospin basis. Engineering a longitudinal gradient of the optical coupling synthesizes an effective pseudo-magnetic field that acts oppositely on these two states. The resulting state-dependent force drives an ultrafast Stern–Gerlach (USG) deflection, spatially separating the wavepacket via its synthetic optical degrees of freedom without relying on the intrinsic magnetic dipole moment.

If the intrinsic momentum spread broadens such that neighboring sidebands overlap, the system crosses over into the anomalous Bragg regime. Here, the interplay between dispersion and Rabi cycling generates a non-classical fishbone interference pattern in Wigner phase space [48]. This complex phase-space topology is physically distinct from the free-drift-induced APINEM limit.

**Vacuum-induced chirping for electron vs material-induced chirping for light.** Free electrons acquire group-velocity dispersion (GVD) during vacuum propagation due to their nonlinear energy–momentum relation, whereas optical pulses propagate dispersion-free. This vacuum GVD becomes dominant at lower velocities, where the longitudinal effective mass enhances wavepacket spreading. Compressing an electron toward the localized limit requires a larger momentum spread, which accelerates free-space expansion. The classical point-particle approximation breaks down if the compressed wavepacket remains broader than the optical wavelength[15]. Under this lens, swift-electron APINEM and slow-electron anomalous Bragg regimes[48] both fall under this generalized highly chirped framework.

During post-interaction free drift, the energy or momentum modulation converts into longitudinal density modulation (bunching) for attosecond electron-beam shaping. In the localized DLA-like limit, a kinematic velocity chirp produces classical bunching through ballistic temporal focusing. In the extended PINEM limit, sideband phase accumulation produces attosecond quantum density bunching through matter-wave interference. (Fig.7)

**Pre- versus post-interaction vacuum dispersion.** Pre- and post-interaction dispersion share the same GVD physical origin but act at different stages of the LEM sequence. During the brief modulation inside a thin LEM, the spatial density remains approximately invariant, restricting the interaction to the momentum-energy phase space. During subsequent free drift, the imprinted momentum distribution is conserved, while vacuum GVD dynamically sculpts the spatial density profile. The distinct physical manifestations of these processes arise strictly from their temporal ordering relative to the optical phase mask.

In cascaded LEM architectures, multiple LEM modules are interleaved with dispersive free-drift sections. Because discrete phase imprinting and continuous vacuum GVD do not commute, these sequences preclude simple analytical approximations and require numerical integration. However, this dynamical complexity unlocks an expansive phase space for tailoring arbitrarily complex wavefunction distributions. Such programmable cascades provide a route toward attosecond beam shaping, structured superradiant emission, and free-electron quantum information processing. We numerically evaluate the modulated momentum spectra to visualize how vacuum dispersion dictates anomalous multiphoton scattering (Fig. 5). For swift electrons (Figs. 5a, d), the interaction embeds the APINEM response within an unbounded multi-level Raman–Nath manifold. For slow electrons, recoil and phase matching confine the dynamics to effective three-level (Figs. 5b, e) or two-level Bragg (Figs. 5c, f) subspaces, restricting the available scattering pathways. Across these regimes, the spectral topology is strictly controlled by the interplay between the initial momentum spread $\sigma_p$ and the accumulated vacuum chirp $\xi t$. Under minimal pre-chirp, the spectra manifest as rigid, vertical sidebands. As the pre-chirp increases, the broadened adjacent components overlap in momentum space. This coherent overlap drives inter-sideband interference, producing oscillatory and branched spectral structures. Transitioning into the strong-coupling regime ($2|g| = 9$, Figs. 5d–f) populates a wider phase space, exacerbating the GVD-induced branching. Ultimately, these numerical phase-space topologies confirm that the APINEM regime is a deterministic quantum interference landscape sculpted by free-space vacuum GVD, rather than a simply smeared PINEM comb.

## Operational taxonomy of light-electron electro-optic modulators

A coherent free electron can be viewed as a high-frequency matter-wave carrier, analogous to the optical carrier in an electro-optic modulator. In this modulator picture, the optical near field plays the role of a drive waveform, while the electron wavefunction carries the encoded phase, energy, momentum, or temporal structure to downstream electron-optical readout. This input-drive-output relation defines the functional scope of the light-electron modulator (LEM). We therefore classify LEMs by the geometry of electron-photon coupling and by the electron degree of freedom that is modulated: longitudinal energy-time, transverse momentum-position, or hybrid phase-space coordinates.

### A. Transverse and longitudinal phase modulation: KDE and PINEM

PINEM and KDE represent two complementary limits of LEM operation: PINEM primarily modulates the longitudinal energy-time degree of freedom, whereas KDE [21] modulates the transverse momentum-position degree of freedom. In KDE, a standing optical wave acts as an intensity grating that imprints a transverse phase modulation on the electron wavefunction. This interaction effectively reverses the traditional roles of light and matter, utilizing the optical field as a coherent, programmable material grating.

When an electron crosses an optical standing wave, the time-averaged ponderomotive interaction produces a periodic transverse phase grating $V_p(x) = V_0 \cos^2(kx)$, where $V_0 = e^2 I/(2m\varepsilon_0 c\omega_L^2)$ parameterizes the potential depth dictated by the peak optical intensity $I$ and laser angular frequency $\omega_L$. The corresponding transverse single-particle Hamiltonian simplifies to $H = \frac{p_x^2}{2m} + V_0 \cos^2(kx)$. By expanding the transverse wavefunction in the plane-wave basis, $\psi(x, t) = \sum_n c_n(t) e^{inkx}$, the temporal evolution of the momentum coefficients is mapped onto a tight-binding lattice. Crucially, the spatial periodicity of this ponderomotive potential dictates that momentum transfer occurs exclusively via even-photon recoil ($\Delta p = \pm 2\hbar k$).

The mode amplitudes evolve according to the coupled-mode equations in energy units:

$$i\hbar \frac{\mathrm{dc_n}}{dt} = \left(\delta_n + \frac{V_0}{2}\right) c_n + \frac{V_0}{4}(c_{n-2} + c_{n+2})$$

where $\delta_n = n^2 \epsilon_r$ specifies the kinetic recoil detuning, with $\epsilon_r = \frac{\hbar k^2}{2m}$ is the single-photon recoil energy. By evaluating the competition between the on-site kinetic detuning $\delta_n$ and the off-diagonal coupling strength $V_0/4$, we can classify the KDE diffraction landscape into the Raman–Nath and Bragg regimes. In the Raman–Nath (thin-grating) limit, the interaction is short enough that recoil detuning can be neglected. The probability of populating the $m$-th diffraction order (corresponding to a net momentum shift of $2m\hbar k$) follows the Bessel-function statistics, $P_{2m}(t) = |c_{2m}(t)|^2 = J_m^2(V_0 t/2\hbar)$ (Fig. 4). This analytical result establishes the central LEM analogy: KDE is the transverse-momentum counterpart of PINEM energy-sideband modulation.

In the complementary Bragg limit, the interaction selectively couples strictly two resonant diffraction orders, driving coherent Rabi oscillations:

$$|c_{+1}(t)|^2 = \cos^2\left(\frac{V_0 t}{4\hbar}\right), \quad |c_{-1}(t)|^2 = \sin^2\left(\frac{V_0 t}{4\hbar}\right)$$

Unlike PINEM, this transverse interaction strictly exchanges transverse momentum while leaving the macroscopic longitudinal electron energy unperturbed. Consequently, the scattering dynamics are completely agnostic to the optical carrier-envelope phase, depending exclusively on the standing-wave intensity envelope and strict phase-matching conditions. The main wavepacket requirement is therefore transverse coherence across the optical grating; the longitudinal duration is less restrictive than in longitudinal energy modulation. Kinematically, evaluating the asymptotic scattered momentum state $\boldsymbol{p} = (p_z', p_x + 2n\hbar k)$ reveals that the wavepacket absorbs and stimulates the emission of optical quanta from the counter-propagating fields, shifting the transverse momentum by discrete intervals of $2n\hbar k$ while the total kinetic energy is conserved in the laboratory frame. In free space, exact energy-momentum conservation strictly dictates the discrete emission angles of these diffracted orders.

At the Hamiltonian level, KDE and PINEM share the same phase-modulation structure, but they address different electron coordinates. However, translating this theoretical isomorphism into operational LEM devices reveals distinct wavepacket constraints, particularly in the ultrafast regime. A representative transverse example is the ultrafast Kapitza–Dirac effect (UKDE). Unlike conventional KDE—which scatters a narrowband incident momentum distribution into discrete, well-resolved sidebands—the UKDE manifests as a periodic momentum-space compression of a broadband spectrum[22].

Within the LEM taxonomy, UKDE is the transverse-momentum analogue of longitudinal APINEM. Just as the longitudinal APINEM necessitates an incident wavepacket with a massive temporal chirp (a longitudinal energy spread vastly exceeding the single-photon quantum $\hbar\omega_L$), the UKDE strictly requires a highly pre-chirped transverse wavepacket. This symmetry shows that dispersion-induced anomalous modulation is not limited to longitudinal PINEM, but also appears in transverse KDE-type modulators.

### B. Linear versus ponderomotive modulation: $\boldsymbol{p} \cdot \boldsymbol{A}$ vs. $\boldsymbol{A}^2$ limits

Starting from the minimal-coupling Hamiltonian, the electron–light interaction separates into two modulation channels: the linear $p \cdot A$ term and the quadratic $A^2$ term. The primary channel is the linear coupling term ($H_I \propto A$), while the secondary channel is the nonlinear ponderomotive term ($H_I \propto A^2$). Typically, for high-energy electrons or moderate optical intensities, the linear channel dictates the interaction dynamics. Conversely, the ponderomotive channel assumes dominance under ultra-intense optical driving or in strictly orthogonal interaction geometries where the linear term vanishes ($p \cdot A \approx 0$).

From a device–engineering perspective, this physical duality provides a direct operational analogue to the nonlinear susceptibility expansion in solid-state electro-optic modulators (EOMs), where the field-dependent refractive index is described by $n = n_0 + \gamma E + \beta E^2$. The linear $p \cdot A$ coupling serves as the free-electron equivalent of the Pockels effect ($\gamma E$), whereas the quadratic $A^2$ scattering identically mirrors the Kerr effect ($\beta E^2$). In conventional EOMs, the Pockels response usually dominates unless symmetry suppresses it or the optical field is sufficiently strong. Similarly, while linear modulation inherently prevails in standard LEM configurations, isolating the ponderomotive regime requires deliberate architectural choices. Thus, choosing a linear or ponderomotive LEM channel becomes a design problem involving optical intensity, near-field symmetry, and electron trajectory.

### C. Travelling-wave versus standing-wave LEM geometries.

The optical-field geometry strongly influences which LEM channel is selected. A traveling-wave geometry provides a nearly uniform cycle-averaged intensity and therefore favors linear, PINEM-type modulation through the longitudinal vector potential. Conversely, a standing-wave geometry synthesizes a stationary spatial interference pattern, generating a periodic optical lattice. This gradient enhances the $A^2$ term and naturally drives ponderomotive, KDE-type modulation. This traveling-wave/standing-wave distinction is a useful engineering rule, but not a strict separation. Through rigorous phase-matching engineering, one can strategically isolate either interaction channel in both geometries.

Traveling-wave LEMs are especially suited for swift electrons and extended interaction lengths. By meticulously engineering the dielectric or photonic structure to synchronize the optical phase velocity with the relativistic electron velocity ($v_e = v_{\mathrm{ph}}$), one achieves sustained co-propagating phase matching. This continuous spatial synchronization is strictly required for broadband modulation, high-order multiphoton embedding, and macroscopic coupling efficiencies.

Standing-wave LEMs are more suitable for slow electrons and strong-recoil regimes. For low-energy beams, strong recoil makes long-distance phase-velocity matching difficult. A standing-wave architecture bypasses this longitudinal constraint; its localized nodal structure accumulates scattering phase abruptly at discrete spatial coordinates, relaxing the requirement for continuous co-propagation. Thus, traveling-wave LEMs favor high-speed, multi-level energy modulation, whereas standing-wave LEMs favor robust transverse control in strong-recoil regimes. Both geometries act as programmable optical phase masks for shaping free-electron wavefunctions.

**D. Transverse and longitudinal LEM channels.**

The geometric orientation of the electron trajectory relative to the driving optical field strictly delineates the macroscopic functionality of the LEM. In longitudinal modulation architectures, the electron co-propagates parallel to the dominant electric-field vector. In transverse LEMs, the optical field redirects the electron into different transverse momentum channels, while the coupling strength is still set by the field component projected along the electron trajectory. In this language, KDE is the archetype of transverse momentum modulation, whereas PINEM is the archetype of longitudinal energy modulation. Nevertheless, this geometrical dichotomy is not immutable; for instance, the coherent beating of two multi-color laser fields can synthesize a longitudinal KDE potential[20,57], proving that the interaction topography can be arbitrarily engineered.

Functionally, transverse LEMs convert optical modulation into angular separation, enabling beam deflection, spatial sorting, and matter-wave beam splitting. Longitudinal LEMs preserve the beam axis but reshape the energy–time structure, enabling energy combs, attosecond streaking, and density bunching. Crucially, these operational modes are governed by distinct kinematic boundaries: transverse modulation dictates rigorous transverse phase matching (transverse momentum conservation) across the extended optical wavefront, a kinematic constraint that is inherently relaxed in purely one-dimensional longitudinal geometries.

**E. Amplitude (AM) versus phase (PM) modulations.**

As in AOMs and EOMs, LEMs can be classified by how the optical drive is mapped onto the electron wavefunction. The relevant device-level outputs include amplitude-like contrast, frequency or energy shifts, phase modulation, and transverse momentum modulation. These regimes are dynamically interconnected; their distinction lies entirely in how the macroscopic optical envelope is inscribed onto the coherent matter wave.

Pure AM must be defined carefully for a single free-electron wavefunction. For an electron initialized in a zero-drift state, a phenomenological factor such as $[1 + m\cos(\omega_m t)]$ multiplying the free-particle state would generally lead to a time-dependent norm unless an explicit normalization or post-selection procedure is specified. An AM-like electron state is therefore better represented as a normalized superposition of the carrier and selected sidebands.:

$$\psi_{\mathrm{AM}}(t) = c_0 e^{-iE_0 t/\hbar} + c_{+1} e^{-i(E_0+\hbar\omega_m)t/\hbar} + c_{-1} e^{-i(E_0-\hbar\omega_m)t/\hbar}$$

with

$$|c_0|^2 + |c_{+1}|^2 + |c_{-1}|^2 = 1$$

This construction defines a three-component coherent state consisting of the unperturbed carrier and two symmetric energy sidebands. In a scattering interpretation, these sidebands correspond to the first gain and loss channels associated with single-quantum exchange. It should not be interpreted as the generic unitary output of a standard LEM. Under minimal coupling, the optical near field primarily imprints phase, producing the Bessel-weighted multiphoton sideband comb characteristic of PINEM. Obtaining an AM-like electron density or a genuinely truncated sideband manifold therefore requires additional operations, such as energy/momentum filtering, engineered mode selectivity, Bragg-type truncation, or free-space ballistic propagation that converts phase or energy modulation into longitudinal density bunching. In this sense, the three-level AM expression is best understood as a controlled spectral ansatz or post-selected state, rather than as the direct, unitary output of standard PINEM phase modulation.

Conversely, frequency and phase modulation fundamentally operate by imprinting a time-dependent geometric or dynamical phase onto the coherent matter wave. Both are therefore naturally described as generalized phase modulation along the electron trajectory. Adopting the identical initial conditions used for the AM derivation, the temporally phase-modulated electron carrier is mathematically expressed as:

$$\psi_{PM}(t) = \varphi e^{-\frac{iE_0 t}{\hbar} - i\phi(t)} = \varphi e^{-\frac{iE_0 t}{\hbar} - im_p \cos\omega_m t}$$

where $\phi(t)$ denotes an arbitrary generalized phase profile. Assuming a canonical harmonic near-field interaction—such as the established modulation within the strongly enhanced optical near-field of a nanotip [11]—this profile takes the form $\phi(t) = m_p \cos(\omega_m t)$, with $m_p = 2|g|$ defining the dimensionless phase-modulation depth (effective coupling parameter).

To map this temporal phase accumulation onto the observable energy spectrum, we apply the Jacobi–Anger expansion to project the wavepacket back into the plane-wave energy eigen basis:

$$\psi_{PM}(t) = \varphi e^{-\frac{iE_0 t}{\hbar}} \sum_{n=-\infty}^{+\infty} (-i)^n J_n(m_p) e^{in\omega_m \mathrm{t}}$$

This rigorous analytical expansion proves that single-frequency harmonic phase modulation naturally spawns an infinite, discrete momentum comb. The resulting spectrum comprises the unperturbed central carrier ($n = 0$) flanked by an unbounded hierarchy of symmetric multiphoton sidebands. Each resonant sideband is separated by the precise energy quantum $\hbar\omega_m$, and its population probability is deterministically governed by the corresponding $n$-th order Bessel function of the first kind, $J_n(m_p)$. Remarkably, this unbounded state-space provides a direct bridge back to the AM regime. In the strict weak-coupling limit ($m_p \ll 1$), the higher-order Bessel modes asymptotically vanish ( $J_n(m_p) \to 0$ for $|n| > 1$), truncating the interaction exclusively to the single-photon exchange domain ($n = \pm 1$).

Beyond monochromatic driving, structured optical waveforms provide additional control knobs for LEM design. Two primary architectural extensions are of particular operational significance: **(i) Optical envelope modulation (CW versus pulsed).** The macroscopic temporal envelope of the driving field strictly dictates the operational mode of the electron beam. Continuous-wave (CW) driving supports steady-state modulation of high-repetition-rate or continuous electron beams. Pulsed driving acts as an ultrafast temporal gate, selecting or reshaping electrons within femtosecond-to-attosecond time windows. Thus, CW LEMs are suited for high-throughput steady-state operation, whereas pulsed LEMs are suited for ultrafast microscopy, streaking, and attosecond bunching. **(ii) Multi-color spectral synthesis.** While single-frequency modulation is kinematically restricted to generating symmetric multiphoton combs or simple frequency shifts, coherently superimposing multiple optical frequencies fundamentally restructures the interaction Hamiltonian. Multi-color or broadband driving permits the Fourier synthesis of arbitrary optical phase masks [58]. This sophisticated spectral control breaks the inherent symmetry of the interaction, enabling asymmetric momentum scattering, precise wavepacket routing, and the deterministic preparation of high-dimensional quantum states.

### F. Phase-space dynamics and Density bunching

In the classical limit, the electron-beam ensemble is described by a phase-space distribution $f(z, p)$[59]. A phase-synchronous LEM, such as a DLA-type near-field modulator, imparts a sinusoidal momentum kick $p = p_0 + \Delta p \sin(kz_0)$ , which converts into longitudinal microbunching during subsequent free drift $z = z_0 + \eta p$, where $\eta$ represents the second-order vacuum dispersion accumulated over the drift time. Assuming an initially uniform electron beam with a Gaussian momentum spread $\sigma_p$, coordinate transformation and Jacobi–Anger expansion yield the classical bunching factor for the $h$-th harmonic:

$$b_f(h) = (-1)^h J_h(hk\Delta p) \exp\left(-\frac{h^2k^2\eta^2\sigma_p^2}{2}\right)$$

This expression contains two competing factors: the deterministic ballistic bunching dictated by the Bessel function, which peaks at the optimum phase-space shear parameter $hk\eta\Delta p \approx 1.84$, and the Gaussian envelope acting as a thermal coherence suppression factor. Substituting the optimum shear condition ( $k\eta \approx 1.2/\Delta p$), the suppression term reduces to $\exp\left[-0.72\left(h\sigma_p/\Delta p\right)^2\right]$ [60,61] , establishing the condition $h\sigma_p \ll \Delta p$ for efficient high-harmonic bunching. This classical bunching factor is an ensemble metric: it describes the statistical phase-space shearing widely used in free-electron lasers and accelerator physics.

Quantum density bunching describes the same phase-to-density conversion at the single-electron wavefunction level. It can be quantified by the expectation value of the spatial harmonic operator, or equivalently by the Wigner distribution $W(z,p)$:

$$b_f(h) = \langle\psi|e^{-ihkz}|\psi\rangle = \int dzdpe^{-ihkz}W(z,p)$$

This quantity is the h-th Fourier component of the single-electron density $\rho(z,t) = \int dp\, W(z,p) = |\psi(z,t)|^2$ (with $\int dz\, \rho = 1$). Assuming a weak modulation regime ($\Delta p \ll p_0$) where the non-recoil velocity approximation $v(p) \approx v_0$ is valid, the corresponding quantum current density evaluates exactly to $J(z) = -ev_0 \int dp\, W(z,p)$. By operating in this normalized Fourier space, the quantum bunching factor $b_f(h)$ elegantly functions as a dimensionless indicator ( $J_h/J_0$) that perfectly captures the attosecond density beat notes induced by multiphoton interference, physically distinguishing it from purely classical particle kinematics.

For a Gaussian incident QEW, the instantaneous PINEM interaction produces a coherent superposition of finite-width sidebands:

$$\psi(p) = \left(\frac{1}{2\pi\sigma_p^2}\right)^{1/4} \sum_n J_n\,(2|g|)e^{-in\phi_0}\, exp\left(-\frac{(p-p_0+n\delta p)^2}{4\sigma_p^2}\right)$$

Transforming this momentum-space comb into the real spatiotemporal domain via Fourier synthesis yields the dynamic spatial wavefunction. Accounting for the intrinsic vacuum dispersion ( $\xi$ ) and the kinematic velocity shift of each sideband ( $\delta v \approx \delta p/m^*$ ), the propagated wavepacket evaluates to:

$$\psi(z,t) = \frac{e^{e^{i(p_0 z - E_0 t)\hbar}}}{\left(2\pi\sigma_{z_0}^2\right)^{1/4}\sqrt{1+i\xi t}} \sum_{n=-\infty}^{+\infty} J_n(2|g|)\, exp\left[-\frac{(z - \mathrm{v}_0 t + n\delta \mathrm{v}t)^2}{4\sigma_{z_0}^2(1+i\xi t)}\right] e^{-i\left(\frac{n\omega_L}{\mathrm{v}_0}\right)(z - \mathrm{v}_0 t - n\delta v t)}$$

where $\sigma_{z0} = \hbar/(2\sigma_p)$ defines the initial spatial coherence length. To fully capture the underlying phase-space topology and directly contrast it with the classical distribution $f(z,p)$, we construct the Wigner quasi-probability distribution. Evaluated in the comoving spatial frame ( $\zeta = z - v_0 t$ ) with respect to the shifted momentum ( $p' = p - p_0$ ), the time-dependent Wigner function is defined by the momentum-space autocorrelation:

$$W(\zeta, p', t) = \frac{1}{2\pi\hbar}\int dq\, \psi^*\left(\mathrm{p}' - \frac{q}{2}\right)\psi(\mathrm{p}' + \frac{q}{2})\, e^{-\frac{i\left[E\left(\mathrm{p}'+\frac{q}{2}\right) - E\left(\mathrm{p}'-\frac{q}{2}\right) - v_0 q\right]t}{\hbar}} e^{\frac{iq\zeta}{\hbar}}$$

While this Wigner formulation strictly preserves the macroscopic normalization condition, $\int d\zeta dp' W(\zeta, p', t) = 1$, analogous to the classical phase-space density, it fundamentally diverges by admitting negative quasi-probability values. Such negativity appears as interference fringes in phase space. These fringes diagnose coherent LEM-driven wavepacket shaping beyond classical phase-space shear[62–66].

Applying the Wigner phase-space formalism, one can rigorously derive the dynamic evolution of the quantum bunching factor during post-interaction free drift. For an electron interacting with an optical near-field, the $n$ -th order coherent density harmonic evaluated after a drift time $t_d$, takes the exact analytical form [63–65]:

$$b_f(n, t_d) = J_n\left[4|g|\sin\left(\frac{\mathrm{n}\delta p^2}{2\hbar\gamma^3 m_0} t_d\right)\right] \exp\left[-\frac{n^2}{2}\left(\frac{t_d}{\gamma^3 m_0}\right)^2 \sigma_k^2 \delta p^2\right] e^{-in\phi_0}$$

, where $\delta p = \hbar\omega_L/v_0$ is the longitudinal momentum spacing associated with one photon-energy exchange, and $\sigma_k = \sigma_p/\hbar$ denotes the intrinsic wave-vector spread of the incident electron wavepacket.. The optical phase $\phi_0$ is retained because it fixes the phase of the density harmonic and enables phase-coherent averaging over many electrons.

The Wigner picture shows that free drift shears the momentum-modulated wavepacket into a longitudinal density modulation. The density contrast is maximized when the drift-induced spatial separation becomes comparable to the optical modulation period. Estimating the maximal sideband-induced velocity modulation as $\Delta v_m \approx \frac{\Delta p_m}{\gamma^3 m_0}$ with $\Delta p_m \approx 2|g|\delta p$, the optimal drift time is approximately:

$$t_{d,\mathrm{opt}} \approx \frac{1}{k_s \Delta v_m} = \frac{\gamma^3 m_0}{k_s \Delta p_m} = \frac{T_L}{2\pi}\frac{\gamma^2 p_0}{\Delta p_m},$$

where $k_s = \omega_L/v_0$, $T = 2\pi/\omega_L$, and $p_0 = \gamma m_0 v_0$. Because the absolute momentum spectrum constitutes a strictly conserved quantity during free-space propagation, purely spectral measurements (e.g., standard energy-loss spectroscopy) are fundamentally blind to this dynamical spatial structuring. Resolving the evolved microbunched topology necessitates time-resolved spatial probing or coherent radiation analysis.

Collectively, the quantum bunching requires the three indispensable pillars: (1) discrete multiphoton sideband generation, (2) dispersion-driven momentum-to-density modulation conversion, and (3) nonclassical phase-space fringes. These interconnected mechanisms form the rigorous theoretical foundation for the exploration on high-harmonic generations and radiation schemes via modulated free electrons [65,67]—a coherent tabletop analogue to high-gain harmonic generation (HGHG) in free-electron lasers[68,69].

**G. Classical versus quantized light modulation**

For coherent laser drives with large photon occupation, the semiclassical phase-modulation picture is usually sufficient. When the drive has nonclassical photon statistics, such as Fock, squeezed, or entangled states, the optical field must be quantized. In that regime, LEM describes not only optical modulation of the electron, but also the transfer of quantum statistics and correlations between photons and free electrons[12–14,70–74].

A compact quantized description treats the electron sidebands and the optical mode as coupled quantum ladders. The joint state evolves under the total Hamiltonian $\hat{H} = \hat{H}_0 + \hat{H}_{\text{ph}} + \hat{H}_I$. We first second-quantize the free-electron field operator by expanding it in a fermionic Bloch-wave basis, $\hat{\psi}_e(z) = \frac{1}{\sqrt{L}} \sum_n \hat{c}_n \, e^{ik_n z}$, where $L$ is the quantization length and the sideband operators $\hat{c}_n$ satisfy standard fermionic anticommutation relations. The index $n$ represents the discrete multiphoton momentum ladder, with sideband wavevectors $k_n = (p_0 + n\delta p)/\hbar$ and $\delta p = \hbar q_z$. Under the small-recoil approximation ($q_z \ll p_0/\hbar$),, the free-electron Hamiltonian evaluates to:

$$\hat{H}_0 = \sum_{n=-\infty}^{+\infty} \varepsilon_n \, \hat{c}_n^\dagger \hat{c}_n,$$

where $\varepsilon_n = E_0 + v_0(\hbar k_n - p_0) + \frac{(\hbar k_n - p_0)^2}{2\gamma^3 m_0}$ defines the quadratic kinetic dispersion.

Concurrently, the mediating electromagnetic near field is quantized as a superposition of bosonic modes. For a single dominant optical mode, the field Hamiltonian is simply to $\hat{H}_{\text{ph}} = \hbar\omega_L \left(\hat{a}^\dagger \hat{a} + \frac{1}{2}\right)$. Enforcing the strict minimal coupling convention for a negatively charged electron ( $q = -e$ ), the perturbative interaction Hamiltonian is governed by the coupling term $\hat{H}_I = \frac{e}{2\gamma m_0} \int dz \hat{\psi}_e^\dagger(z) \left[\hat{A}_z(z)\hat{p}_z + \hat{p}_z \hat{A}_z(z)\right] \hat{\psi}_e(z)$. In the small-recoil limit ($\hat{p}_z \approx p_0$), spatial integration and the phase-matched rotating-wave approximation simplify the interaction to:

$$\hat{H}_I = \hbar \sum_n \left(g_q \hat{c}_{n+1}^\dagger \hat{c}_n \hat{a} + g_q^* \hat{c}_n^\dagger \hat{c}_{n+1} \hat{a}^\dagger\right).$$

Combining these elements yields the definitive second-quantized LEM Hamiltonian:

$$\hat{H} = \sum_{n=-\infty}^{+\infty} \varepsilon_n \, \hat{c}_n^\dagger \hat{c}_n + \hbar\omega_L \left(\hat{a}^\dagger \hat{a} + \frac{1}{2}\right) + \hbar \sum_n \left(g_q \hat{c}_n^\dagger \hat{c}_{n-1} \hat{a} + g_q^* \, \hat{c}_n^\dagger \hat{c}_{n+1} \hat{a}^\dagger\right)$$

where the effective coupling coefficient $g$ seamlessly absorbs the localized nanophotonic mode profile ($\widetilde{E_z}$), the macroscopic box-quantization volume, and the kinematic pre-factors.

This rigorous QED formulation provides the universal foundation for exploring free-electron quantum optics in plane-wave limit. It natively recovers classical wavepacket diffraction, such as Raman–Nath random walks, while explicitly incorporating the dynamic quantum statistics of the driving field. Extensions that include dispersive material modes, strong optical nonlinearities, or environmental decoherence belong to a broader macroscopic-QED description, but are beyond the minimal closed-system LEM model considered here.

A rigorous examination of the exact kinematic dispersion reveals that the quadratic dispersion term, $n^2\epsilon = \frac{n^2\delta p^2}{2\gamma^3 m_0}$ , effectively acts as a confining momentum-space potential. Because this recoil-induced detuning scales quadratically with sideband order $n$, high-order sideband transitions are suppressed, especially for slow electrons where recoil and dispersion are large [45]. Depending on the targeted interaction regime, this dispersion barrier allows for the adiabatic elimination of highly detuned states, effectively truncating the infinite-dimensional PINEM Hilbert space into closed three-level ( $n \in \{-1, 0, 1\}$ ) or two-level subspaces.

Under symmetric Bragg diffraction conditions, the multiphoton ladder is rigorously truncated into an effective two-level manifold spanned by the resonant momentum states $n = \pm 1/2$. Within this isolated subspace, the system is identically described by the free-electron Jaynes–Cummings Hamiltonian:

$$\begin{aligned}\widehat{H} &= \frac{\hbar v_0 q_z}{2}\left(c_{\frac{1}{2}}^\dagger c_{\frac{1}{2}} - c_{-\frac{1}{2}}^\dagger c_{-\frac{1}{2}}\right) + \hbar\omega_L \hat{a}^\dagger \hat{a} + \hbar\left(g_q c_{\frac{1}{2}}^\dagger c_{-\frac{1}{2}} \hat{a} + g_q^* c_{-\frac{1}{2}}^\dagger c_{\frac{1}{2}} \hat{a}^\dagger\right) \\ &= \hbar v_0 q_z \hat{\tau}_z + \hbar\omega_L \hat{a}^\dagger \hat{a} + \hbar\left( g_q \hat{\tau}_+ \hat{a} + g_q^* \hat{\tau}_- \hat{a}^\dagger\right)\end{aligned} \tag{2}$$

where we define the pseudospin operators $\hat{\tau}_z = \frac{1}{2}\left(\hat{c}_{1/2}^\dagger \hat{c}_{1/2} - \hat{c}_{-1/2}^\dagger \hat{c}_{-1/2}\right)$, $\hat{\tau}_+ = \hat{c}_{1/2}^\dagger \hat{c}_{-1/2}$, and $\hat{\tau}_- = \hat{c}_{-1/2}^\dagger \hat{c}_{1/2}$ .Constraining the translational degrees of freedom to this synthetic two-level pseudospin natively establishes a free-electron qubit. Governed by the Jaynes–Cummings Hamiltonian, this architecture drives genuine free-electron Rabi oscillations, enabling the coherent, light-driven manipulation of free-electron qubits in ultrafast systems[45–47]. While alternative spatial or mode-based design schemes for free-electron qubits exist[55,56], they inherently lack the underlying symmetry required to support these pristine, pure Rabi dynamics.

Crucially, this intrinsic two-level system—constructed entirely from the free electron's quantized spatial momentum states—must be strictly distinguished from free-electron–bound-electron resonant interaction (FEBERI) architectures [28,31]. In FEBERI, the required two-level nonlinearity originates externally from the Coulomb scattering between the passing electron and a discrete atomic-like bound state. Conversely, the free-electron Jaynes–Cummings model strategically exploits the fundamental vacuum dispersion of the electron wavepacket itself to synthesize a standalone, highly programmable quantum logic basis.

**H. Reciprocal modulations: shaping light by modulated electron.**

Once the interaction Hamiltonian is mapped onto the second-quantized field operators, the coupled dynamics are evaluated via the Heisenberg equations of motion, $i\hbar\,\partial_t\hat{O} = [\hat{O},\hat{H}]$. By evaluating the commutators for the electron and photon modes, we obtain the coupled equations:

$$i\hbar\frac{\partial\hat{c}_n}{\partial t} = \varepsilon_n\hat{c}_n + \hbar g_q\hat{c}_{n-1}\hat{a} + \hbar g_q^*\hat{c}_{n+1}\hat{a}^\dagger,$$
$$i\hbar\frac{\partial\hat{a}}{\partial t} = \hbar\left(\omega_L - i\frac{\kappa}{2}\right)\hat{a} + \hbar g_q^*\sum_n \hat{c}_n^\dagger\,\hat{c}_{n+1},$$

where a phenomenological cavity decay rate $\kappa$ has been introduced to the optical mode to account for finite cavity quality factors.

This coupled system rigorously defines the paradigm of *reciprocal modulation*. The first equation governs the canonical PINEM process—the shaping of the electron matter wave by the quantized optical field. A step along the electron ladder is accompanied by photon annihilation or creation in the optical mode. Conversely, the second equation governs the reciprocal back-action: the shaping of the optical field by the modulated electron. This operator equation is the exact quantum analogue of the classical macroscopic Maxwell wave equation, $\nabla^2\boldsymbol{A} - \frac{1}{c^2}\partial_t^2\boldsymbol{A} = -\mu_0\boldsymbol{J}$. The operator $\sum_n \hat{c}_n^\dagger\hat{c}_{n+1}$ is the quantum analogue of a bunched current harmonic, providing the source term for photon emission or amplification.

This reciprocal architecture maps identically onto the quantum free-electron laser (QFEL) framework [75]. Within this formalism, the macroscopic radiation properties and kinematic limits of the interaction are elegantly captured by two fundamental expectation values: the generalized quantum bunching factor $b_h = \sum_n c_n c_{n+h}^*$, and the quantum acceleration gradient (mean energy exchange) $\langle\Delta E\rangle = \sum_n n\hbar\omega_{\mathrm{L}}|c_n|^2$.

Reciprocal LEM therefore positions free electrons as active quantum interfaces rather than passive probes. It establishes an independent, unifying platform capable of bridging disparate quantum systems. In a longer-term vision, such interfaces could link cavity photons, material or atomic qubits, and ultrafast free-electron sidebands in a coherent feedback loop. Realizing this ultimate convergence—a motivation pioneered in early QFEL literature [75–78]—necessitates the ongoing development of a comprehensive, wavepacket-based macroscopic QED framework to capture the complete operational lifecycle of LEM architectures.

Practical quantum LEMs must go beyond closed-system unitary evolution and include loss and decoherence. This can be formulated through a joint electron-photon density matrix $\hat{\rho}$ governed by a Lindblad master equation:

$$\frac{d\hat{\rho}}{dt} = -\frac{i}{\hbar}[\hat{H}, \hat{\rho}] + \sum_i \gamma_i \left(\hat{L}_i \hat{\rho} \hat{L}_i^\dagger - \frac{1}{2}\{\hat{L}_i^\dagger \hat{L}_i, \hat{\rho}\}\right)$$

In this non-unitary formalism, the Lindbladian jump operators $\hat{L}_i$ capture the irreversible dissipation channels inherent to real-world architectures. Such modeling is essential for entanglement-preserving devices, but it substantially increases computational complexity. We therefore treat open-system free-electron quantum optics as a future extension rather than part of the minimal LEM framework developed here.

**Material platforms and structural architectures for light-electron modulators**

The practical implementation of LEM is constrained by material loss, optical damage threshold, nanofabrication tolerance, electron-beam access, and phase-matching stability. These constraints determine whether an optical near field can support localized scattering, extended energy exchange, transverse diffraction, or resonant wavepacket modulation. We therefore group LEM platforms into three implementation classes: localized metallic emitters and nanotips, extended phase-matched structures such as gratings, prisms, and DLA elements, and integrated dielectric photonic structures such as photonic-crystal slabs and microresonators.

Early PINEM experiments often used plasmonic metallic nanostructures because they provide strong local near-field enhancement[9]. Their main limitations are Ohmic loss, heating, and relatively low optical damage thresholds. These constraints motivate the use of low-loss dielectric and nanophotonic platforms, which can sustain higher optical intensities, support longer interaction length, and enable resonant or phase-matched coupling.

**A. Metallic nanotips: Emission and Localized Scattering**

Metallic nanotips provide the most localized implementation of LEM. They can act both as laser-triggered electron emitters and as nanoscale near-field scatterers [79,80]. When the photoemission laser and the near-field drive are phase-locked, the optical phases can be imprinted at or near the moment of electron emission, enabling internally modulated electron sources[81]. Sharp tungsten or gold tips concentrate optical fields at nanometer-scale apexes, allowing subcycle gating[82] and attosecond-scale electron emission[83–85]. Their strength is extreme field localization; their main limitations are low charge per pulse, sensitivity to heating, and limited interaction length.

For nanotip-based LEMs, the key engineering trade-off is between coherence, brightness, and charge per pulse. Downstream applications such as ultrafast low-energy electron diffraction require high transverse coherence, while scalable modulation requires higher current and stable emission. $LaB_6$ nanotips, microfabricated metal tip arrays, and optimized laser polarization provide routes to improve brightness and flux while preserving useful coherence [86–91]. These developments make nanotips valuable as compact coherent sources, but their localized interaction volume makes them less suited to long-range phase-synchronous acceleration.

**B. Extended phase-matched structures**

Extended structures are used when LEM requires coherent interaction over distances longer than the near-field region of a nanotip. Periodic gratings can supply the reciprocal momentum needed to phase match the optical near field to a swift electron, as in inverse Smith–Purcell and DLA-type interactions[92] Depending on field symmetry and injection phase, the same extended geometry can produce PINEM-like multiphoton sidebands, spectral shearing, or net DLA-like energy transfer. Thus, gratings and DLA elements implement the traveling-wave, phase-synchronous branch of the LEM taxonomy.[93].

Early DLA structures used metallic or plasmonic elements to obtain large near-field gradients, including keV-scale modulation bandwidths and sub-GeV/m acceleration gradients[39]. For higher optical powers and longer devices, however, Ohmic loss, thermal damage, and small vacuum apertures become limiting. All-dielectric platforms therefore provide a more scalable route for DLA-type LEMs. Silicon is attractive for infrared integrated photonics, fused silica offers high damage thresholds in visible and near-infrared operation, and diamond is promising for high-power long-wavelength driving.[94]. These materials support stable, phase-synchronous acceleration and connection LEM design to chip-scale accelerator architectures[6].

Periodic nanostructuring is not the only route to phase matching. A dielectric prism can provide Cherenkov-type phase matching when the incident angle is chosen so that the evanescent-wave phase velocity matches the electron velocity[95]. This bulk-optical approach avoids complex nanofabrication and can extend the interaction length to hundreds of micrometers, producing keV-scale multiphoton sideband broadening. In the LEM framework, prism coupling is important because it shows that phase-synchronous modulation can be achieved through bulk dispersion engineering as well as through periodic gratings.

**C. Integrated dielectric Photonics and microresonators**

Integrated dielectric resonators address the optical-power limitation of single-pass LEM geometries. High-Q microcavities and ring resonators store optical energy and increase the effective interaction length, enabling strong modulation with continuous-wave or low-power optical driving. Silicon nitride microrings, for example, have enabled strong PINEM modulation and large sideband combs with waveguide-coupled powers in the microwatt-to-milliwatt range[96].

In the LEM design framework, resonators provide a compact route to low-threshold phase modulation and on-chip electron wavefunction shaping.

Integrated photonics also enables dispersion engineering for slow-electron modulation. Periodic waveguides and photonic-crystal structures can reduce the optical phase velocity and improve matching to low-energy electrons, including near-infrared coupling to tens-of-keV beams[97]. At the same time, nanocavities can reduce the optical mode volume and enhance the local field [98]. Together, engineered phase velocity and small mode volume provide the main design levers for increasing coupling strength in slow-electron LEMs.

High-$Q$ resonators and small mode volumes also provide a path toward quantized-field LEMs. In this regime, the optical field cannot always be treated as a prescribed classical drive; cavity photon statistics, electron–photon correlations, and back-action can become measurable. Recent cavity-mediated experiments have demonstrated electron–photon pair generation and related quantum optical signatures[12,99]. while microresonator-based electron spectroscopy can probe intracavity nonlinear states such as dissipative Kerr solitons [100]. These examples motivate treating integrated resonators as candidate interfaces between free-electron wavepackets and quantized optical modes[101].

These platforms occupy complementary regions of the LEM design space. Metallic nanotips maximize field localization and timing control; gratings, prisms, and DLA structures provide extended phase-synchronous coupling; and dielectric resonators reduce optical power requirements through energy storage and mode confinement. The choice of platform should therefore be matched to the target observable, whether localized emission timing, energy sideband generation, net acceleration, transverse diffraction, or quantum optical back-action.

Emerging polaritonic and reconfigurable platforms extend this design space. Graphene, hexagonal boron nitride (hBN), and related polaritonic media can confine optical fields to deeply subwavelength scales, potentially increasing light–electron coupling in nanoscale geometries [102–104]. Active electro-optic or thermo-optic tuning could further allow dynamic control of coupling strength, phase matching, or electron routing in integrated LEM circuits [27,105]. These directions are promising, but in the present framework they are best viewed as future implementation routes rather than required components of the core LEM model.

**Application directions enabled by LEM architectures**

The LEM framework connects light–electron interaction physics to functional modules for electron microscopy, accelerators, and quantum electron optics. Depending on where the modulator is placed and which degree of freedom is addressed, LEMs can shape electron energy spectra, temporal density, transverse wavefronts, or electron–photon quantum correlations. The common operating principle is optical-cycle-level control of the electron wavepacket, which can be used either before the sample for state preparation or after the sample for ultrafast readout.

**Longitudinal phase-space control.** Longitudinal LEMs primarily reshape the electron energy spectrum and arrival-time distribution while leaving the mean transverse trajectory unchanged. This enables three useful functions: coherent waveform modulation through PINEM-like sideband generation, discrete frequency shifting through Bragg-type population transfer, and macroscopic energy shifting through DLA-like net acceleration or deceleration.

In the Raman–Nath limit, PINEM generates a symmetric Bessel-weighted energy comb and can shape the temporal density of the electron wavepacket after drift. In the Bragg or phase-matched limit, selected sidebands can be preferentially populated, enabling discrete energy translation or Rabi-like exchange between momentum–energy states. In the localized-wavepacket limit, the same longitudinal modulation appears as DLA-like acceleration or velocity chirping. These regimes provide a compact functional ladder from waveform shaping to energy shifting.

Notably, breaking spectral symmetry is central to unidirectional electron-energy control. By extending the interaction length at moderate field strength, the dynamics can transition from the Raman–Nath regime to a Bragg-type regime, where engineered detuning or an anharmonic secondary potential truncates the multiphoton ladder to an effective two-level subsystem. The resulting free-electron Rabi oscillations enable efficient discrete energy translation. In the macroscopic limit of sustained net photon exchange, this becomes an energy shifter: point-like electrons undergo classical acceleration, whereas coherent matter waves acquire a unidirectional spectral shear analogous to frequency shearing in nonlinear optics.

**Transverse wavefront control and beam shaping.** Transverse LEMs use optical phase gradients or ponderomotive intensity gratings to redistribute the electron wavefunction in angle or transverse momentum. Kapitza–Dirac diffraction provides the basic example: a standing optical wave acts as a transverse phase grating and produces coherent angular orders. This functionality can be used for ultrafast beam deflection, coherent beam splitting, transverse wavefront shaping, and spatial mode sorting.

Structured optical fields can extend transverse LEMs from simple diffraction to programmable phase-plate operation. Spatially tailored optical profiles may imprint lensing, steering, or higher-order phase patterns on the electron beam, allowing dynamic control on femtosecond timescales[22]. In practice, this approach is best viewed as a reconfigurable complement to static magnetic or electrostatic electron optics, rather than a complete replacement for conventional aberration correction.

**Electron-Photon Quantum Interfaces.** In the quantized-field regime, LEMs can function as an interface between free-electron sidebands and optical photon states. When a small number of Bragg-coupled electron momentum states is isolated, the interaction can approach Jaynes–Cummings- or quantum-Rabi-type dynamics. This provides a route to coherent electron–photon exchange, sideband-state control, and conditional quantum-state preparation.

Recent experiments on cavity-mediated electron–photon pairs, electron–photon correlations, and post-selected photonic states show that free electrons can participate in quantum optical processes beyond semiclassical phase modulation [99,106,107]. For the present LEM framework, these results are most useful as evidence that sideband-resolved electron measurements can mediate or diagnose quantum states of confined optical modes[101].

A scalable quantum LEMs requires the single-photon coupling rate, optical loss, electron decoherence, timing jitter, and readout fidelity to be treated together. The relevant design target is not only large coupling rate $g_q$, but sufficient cooperativity relative to cavity decay and electron dephasing [27,73]. This requirement links the quantum-interface application directly to the material choices discussed above: small mode volume, high $Q$, phase matching, and stable electron access.

**Column integration and Functional placement.** When placed near the electron source, an LEM can act as a preprocessing element for wavepacket preparation. A longitudinal phase or velocity modulation followed by free drift can compress a continuous or long electron pulse into a train of short density peaks. This source-end configuration is useful for optical-cycle timing, attosecond pulse synthesis, and stroboscopic electron probes, provided that phase stability and space-charge effects are controlled.

When placed after the sample, an LEM can convert temporal information into an energy or momentum signature before detection. In pump–probe measurements, a phase-locked post-sample modulation can act as an ultrafast streaking or demodulation stage, allowing energy-filtered or momentum-resolved detection to recover timing information even when the detector integrates slowly. This detector-end configuration is therefore suited to temporal gating, lock-in-style readout, and ultrafast spectroscopy.

Downstream LEMs can also be used for electron wavepacket characterization. Spectral shearing methods such as FESSI use controlled energy–time shear to retrieve phase information and reconstruct the electron wavepacket or its Wigner representation[43]. This makes detector-end LEMs useful not only for temporal gating, but also for diagnosing the coherence and phase structure produced by upstream modulation.

**Design challenges and operating criteria of LEM devices**

Scalable LEM operation requires co-optimization of optical confinement, phase-matched coupling, electron phase-space quality, and dissipative channels. The closed-loop Maxwell–Schrödinger workflow in Fig. 10 provides a practical design framework: optical near fields determine the modulation Hamiltonian, electron propagation sets the phase-space response, and loss or decoherence defines the usable operating window.

**Maximizing coherent phase-synchronous coupling.** The relevant figure of merit for LEMs is not peak near-field intensity alone, but the accumulated coupling that remains phase matched along the electron trajectory. Device design should therefore maximize useful longitudinal or transverse coupling while minimizing parasitic kicks, optical loss, heating, and emittance growth. DLA structures rely on long-range phase synchronism, nanotips provide strong but localized coupling, and high-$Q$ cavities enhance the effective interaction length at the cost of tighter phase and thermal stability requirements.

The usable interaction length is the distance over which the electron and optical field remain phase coherent, not simply the physical length of the device. Phase-matching strategies such as dispersion-engineered waveguides, two-color driving, and pulse-front tilt should be evaluated by how much coherent interaction length, modulation depth, and usable electron throughput they provide [108]. This shifts LEM optimization from static field enhancement toward phase stability, coupling accumulation, and coherence-preserving transport.

**Preserving spatiotemporal coherence.** LEM performance depends on maintaining stable phase relationships among the generated energy, momentum, and temporal components of the electron wavepacket. The main coherence limits are source energy spread, emittance, carrier-envelope phase jitter, inelastic scattering, timing drift, and uncontrolled vacuum dispersion. A practical LEM design should therefore specify a coherence budget from source to detector, including phase-locked optical delivery, stable electron transport, and readout methods such as FESSI, attosecond streaking, or event-based EELS when feedback stabilization is required.

**Managing space-charge effects and phase-space degradation.** At macroscopic beam currents, Coulombic space-charge repulsion severely broadens the longitudinal phase space, limiting the usable cascade depth of multidimensional LEM systems. The classical bunching pathway [Fig. 7(a–c)] clearly illustrates that if the delicate inter-sideband phase relationships are degraded by space-charge-induced emittance growth, subsequent downstream modulation stages become dynamically unpredictable. Mitigating these multi-particle decoherence mechanisms requires strict operational compromises: utilizing low-charge-per-pulse beams at megahertz repetition rates, implementing early-stage relativistic acceleration for rapid beam expansion, and deploying active chirp compensation alongside event-based deconvolution at the detector. Reports of LEM performance should therefore state the charge per pulse, repetition rate, compression factor, emittance growth, and timing stability together with the optical coupling strength.

**Co-designing nanophotonics for bidirectional coupling.** LEM devices cannot be optimized by optical field confinement alone. The optical mode, electron trajectory, vacuum channel, fabrication tolerance, laser-damage threshold, thermal load, and readout geometry must be designed together. For quantized-field operation, the relevant comparison is between the single-photon coupling rate $g_q$ and the optical, electronic, and environmental decoherence rates. A useful next step for the field is to report common figures of merit, including coupling strength, synchronous interaction length, phase tolerance, damage threshold, beam transmission, and coherence preservation.

**Thermal and radiation-induced dissipation**. Optical absorption can heat nanophotonic structures, shift resonances, and introduce phase noise, especially in high-$Q$ cavities. Electron irradiation can also charge dielectric surfaces, alter near-field profiles, and promote contamination or carbon deposition in practical vacuum environments. These effects reduce coupling stability and coherence lifetime. They should be included in the device operating window through limits on optical power, electron dose, thermal drift, resonance stability, and vacuum cleanliness.

**On-chip integration and modularity.** Integrated nanophotonic platforms can reduce optical phase drift, lower drive power, and make cascaded LEM stages more reproducible. Waveguide-coupled structures route optical fields to the electron channel, resonators enhance field buildup, and group-velocity engineering supports phase matching over longer distances[93,96,108,109]. At the system level, electro-optic tuning may allow sequential pre-modulation, acceleration, and post-filtering stages to be phase controlled on a shared platform. Standardized optical–electron interfaces would make such modules easier to compare, reproduce, and eventually integrate into electron microscopes or accelerators.

**Engineering quantum LEM.** Quantum LEMs operation adds requirements on state encoding, coherent control, and joint electron–photon readout. Candidate encodings include energy-bin, momentum-bin, time-bin, or dual-rail electron states, but each requires a corresponding preparation, manipulation, and measurement protocol. Rather than presenting quantum LEM as a near-term processor, the present framework identifies the required ingredients: high-$Q$ and small-mode-volume optical structures, low-charge coherent electron pulses, strong timing synchronization, high-fidelity sideband readout, and feedback-compatible device metrics.

Overall, scalable LEM devices will require a shift from demonstrating isolated sideband generation to reporting reproducible operating windows. The most useful device-level metrics are accumulated coupling strength, synchronous interaction length, phase and timing tolerance, electron throughput, emittance preservation, optical damage margin, and coherence lifetime. These metrics connect the theoretical LEM framework to practical device comparison and future integration.

## Summary and outlook

In this work, we developed a theoretical and device-level framework for light–electron modulation (LEM). The framework treats the free-electron wavepacket as the matter-wave carrier and the optical near-field as a classical or quantized modulation drive. Within this picture, longitudinal energy modulation, transverse momentum diffraction, PINEM, DLA-like acceleration, Kapitza–Dirac diffraction, APINEM, and Bragg-regime dynamics are organized as related limits of light–electron phase modulation. This formulation connects the interaction Hamiltonian to practical readout observables. A central conclusion is that LEM performance should be evaluated by accumulated phase-synchronous coupling and coherence preservation, rather than by peak near-field intensity alone.

The same framework also identifies the main engineering constraints for practical devices. Scalable LEM operation requires phase-matched optical confinement, stable electron phase space, controlled optical and electron-induced dissipation, and preservation of sideband coherence through the full beamline. For this reason, future device comparisons should report not only sideband generation or energy gain, but also coupling efficiency, synchronous interaction length, CEP and timing tolerance, electron throughput, emittance preservation, optical damage margin, and coherence lifetime. These quantities provide a more reproducible basis for comparing nanotips, gratings, DLA structures, prisms, resonators, and integrated photonic platforms.

Looking forward, integrated nanophotonics, high- $Q$ resonators, dispersion-engineered waveguides, and reconfigurable optical circuits provide promising routes for improving LEM stability and scalability. In parallel, quantized-field implementations may enable controlled electron–photon correlations, cavity-mediated interactions, and sideband-based quantum-state preparation when coupling, loss, timing jitter, and readout fidelity are jointly optimized. These directions suggest that LEMs can evolve from isolated light–electron scattering experiments into modular components for ultrafast electron microscopy, chip-scale acceleration, attosecond beam control, and free-electron quantum optics. The framework developed here provides a basis for that progression by linking microscopic interaction physics to regime maps, device architectures, and measurable operating criteria.

**Acknowledgments**

Y.P. acknowledges support from the National Natural Science Foundation of China (Grant No. 2023X0201-417-03) and the start-up fund from ShanghaiTech University.

Correspondence and requests for materials should be addressed to Y.P. (yiming.pan@shanghaitech.edu.cn).

**Data and code availability**

The code used for the simulations and analysis will be made publicly available in an open repository upon publication. Additional data supporting the findings of this study are available from the corresponding author upon reasonable request.

**Conflict of interest**

The authors declare no competing interests.

**Figures:**

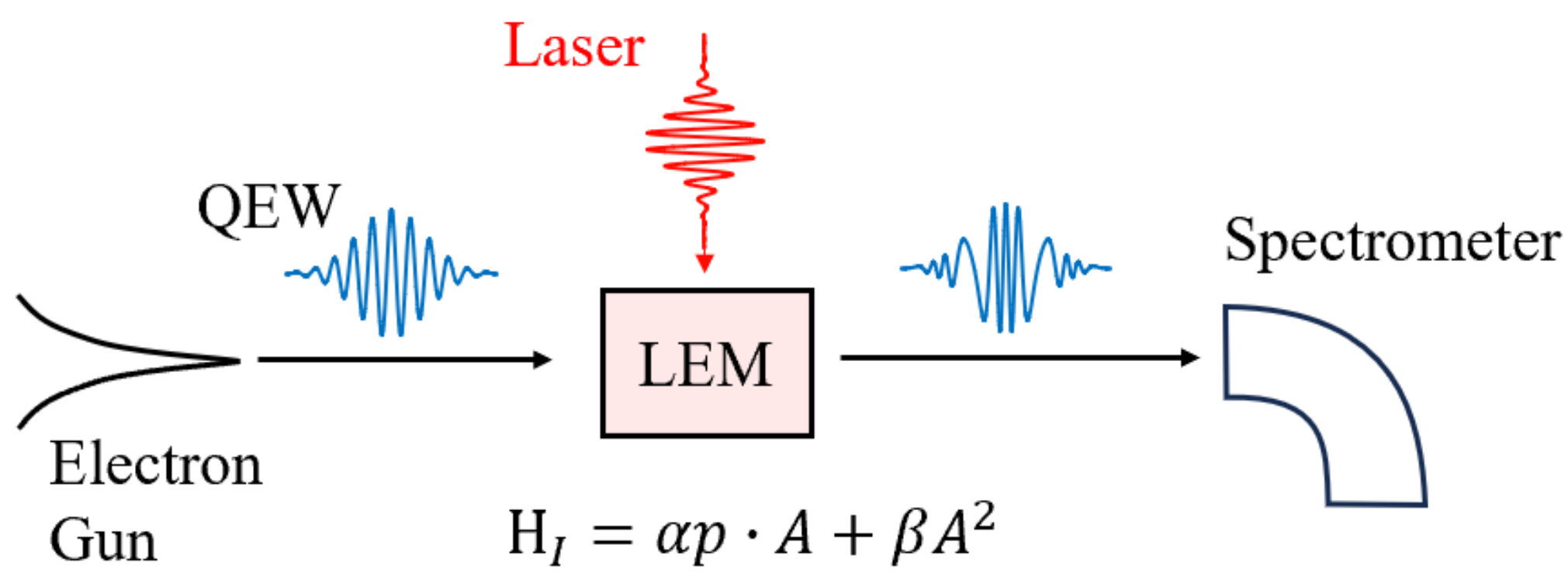

**Figure 1 |** Schematic architecture of a light–electron modulator (LEM). An electron source prepares an incident quantum electron wavepacket (QEW), which interacts with a synchronized optical near field in the LEM region and is analyzed by a downstream spectrometer. In the model developed here, the interaction Hamiltonian $H_I = \alpha p \cdot A + \beta A^2$ contains two modulation channels: the linear $p \cdot A$ coupling and the ponderomotive $A^2$ coupling.

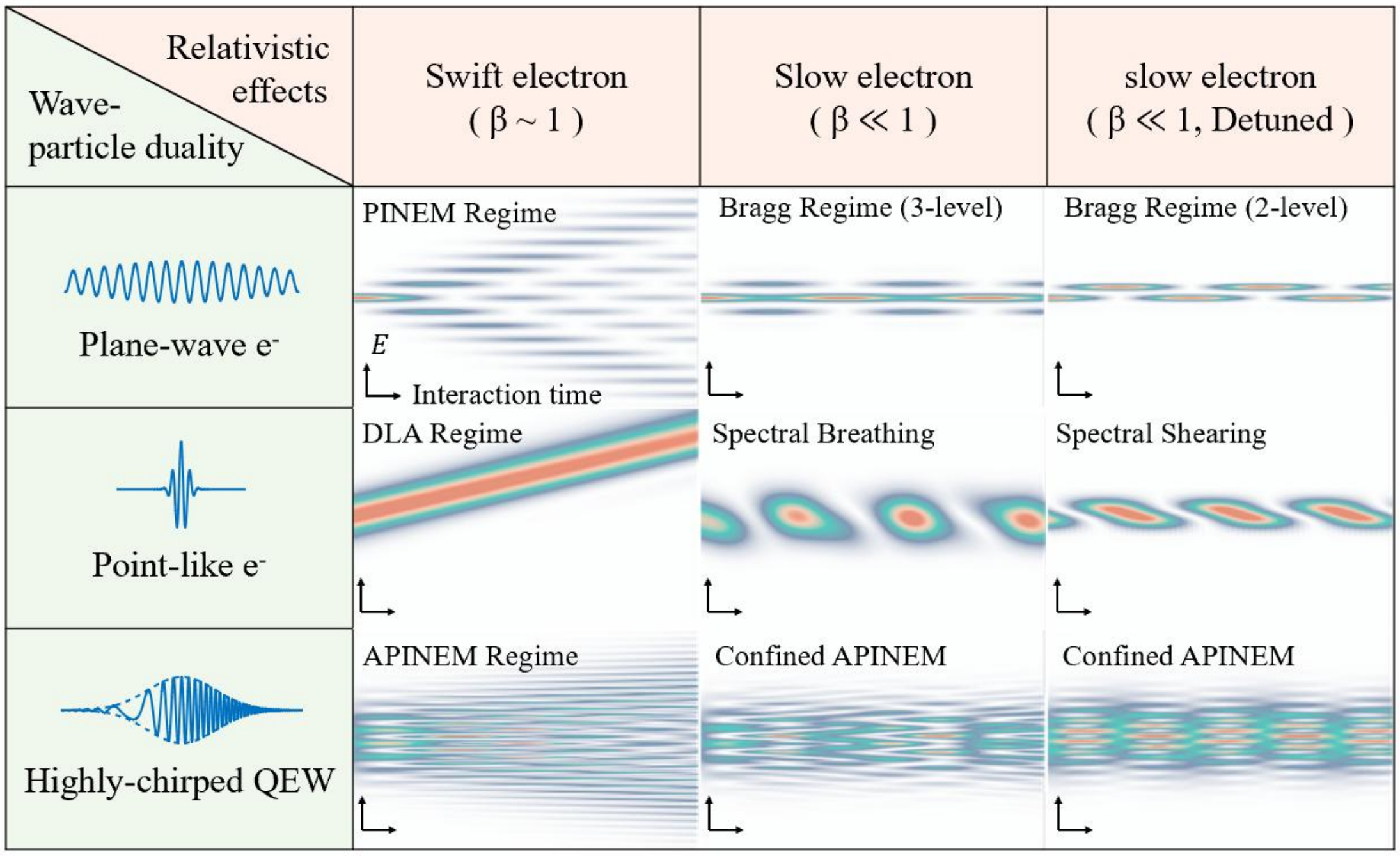

**Figure 2 |** Regime map of light–electron modulation. The map classifies LEM dynamics using two control axes: the longitudinal character of the incident electron wavepacket and the kinematic phase-matching condition. Along the wavepacket axis, the electron evolves from an extended plane-wave-like QEW to a localized wavepacket and then to a highly chirped QEW after free-space dispersion. Along the kinematic axis, swift electrons correspond to Raman–Nath-type modulation, whereas slow or detuned electrons enter Bragg-type regimes where recoil and phase mismatch become important. The resulting parameter space connects PINEM, DLA-like acceleration, spectral breathing, spectral shearing, and APINEM as limiting responses of the same modulation model.

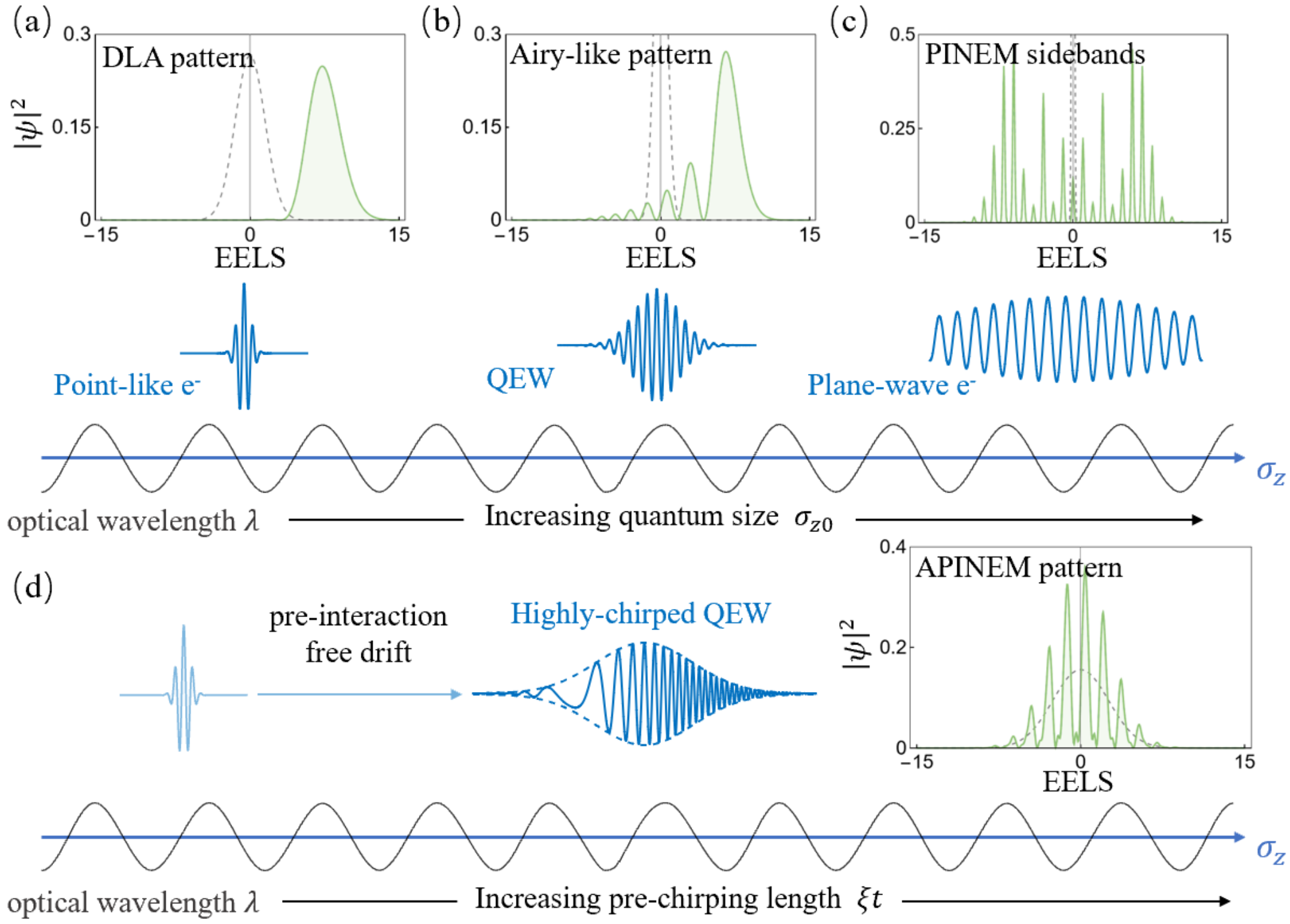

**Figure 3.** Wavepacket-controlled spectral crossover between DLA-like modulation, PINEM, and APINEM. (a–c) Increasing the intrinsic longitudinal wavepacket size $\sigma_{z0}$ changes the spectrum from a continuous DLA-like energy shift to an intermediate Airy-like pattern and finally to a discrete PINEM sideband comb. (d) A localized wavepacket with $\sigma_{z0} \ll \lambda$ can expand during pre-interaction free drift and acquire a longitudinal chirp. This chirped-QEW input produces an APINEM spectrum with periodic spectral bunching on a broad momentum envelope, distinct from both the localized DLA-like limit and the extended PINEM limit.

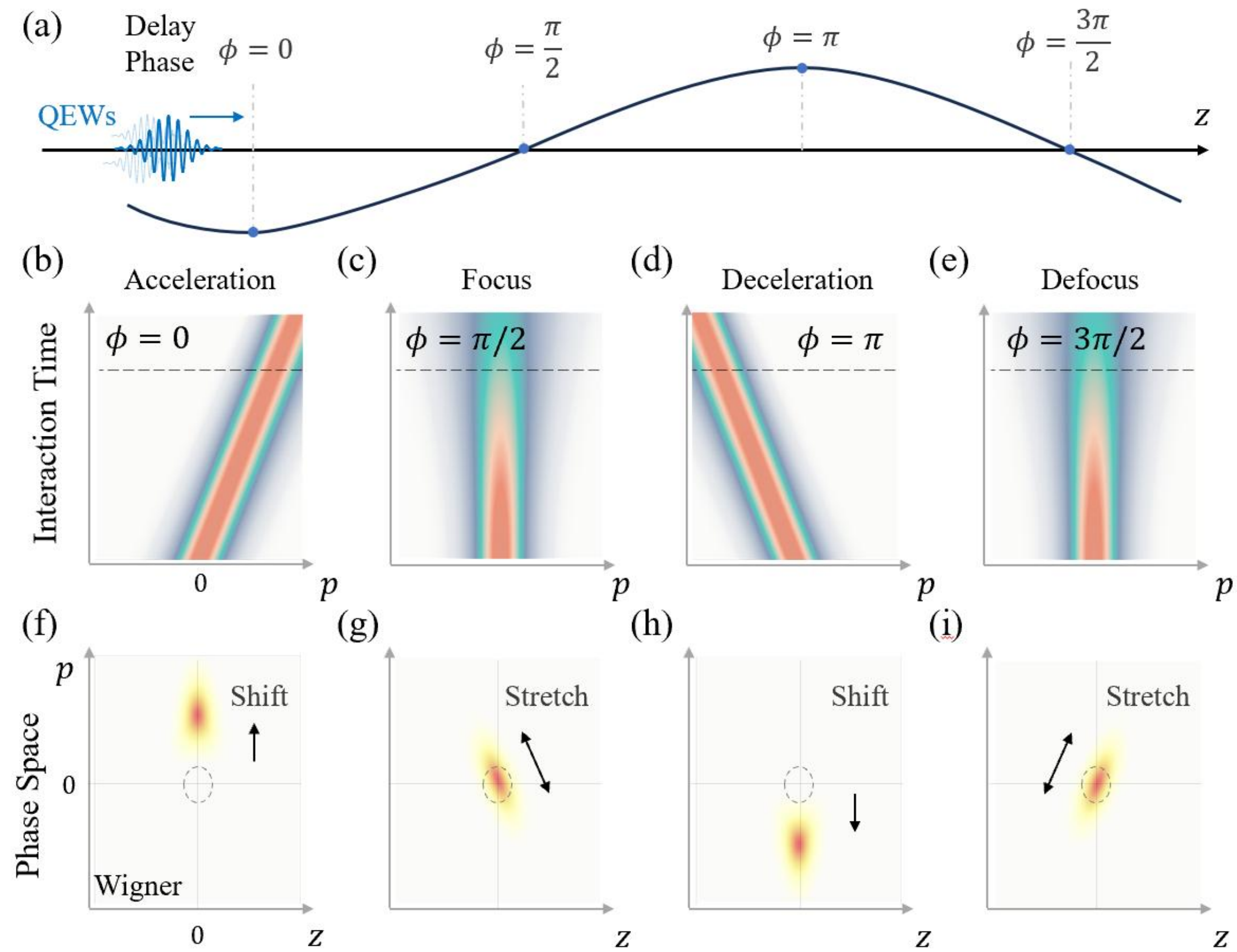

**Figure 4 |** Phase-dependent longitudinal modulation in the localized-wavepacket limit. (a) A localized QEW samples a delay phase $\phi$ of the driving optical field. (b,f) Injection near $\phi = 0$ produces a positive momentum shift corresponding to acceleration, whereas (d,h) injection near $\phi = \pi$ produces a negative momentum shift corresponding to deceleration. (c,g) Injection near $\phi = \pi/2$ produces a velocity chirp with little net energy shift and can focus the wavepacket after drift; (e,i) injection near $\phi = 3\pi/2$ produces the opposite chirp and defocuses the longitudinal wavepacket. The middle panels show spectral evolution, and the lower panels show the corresponding Wigner phase-space distributions.

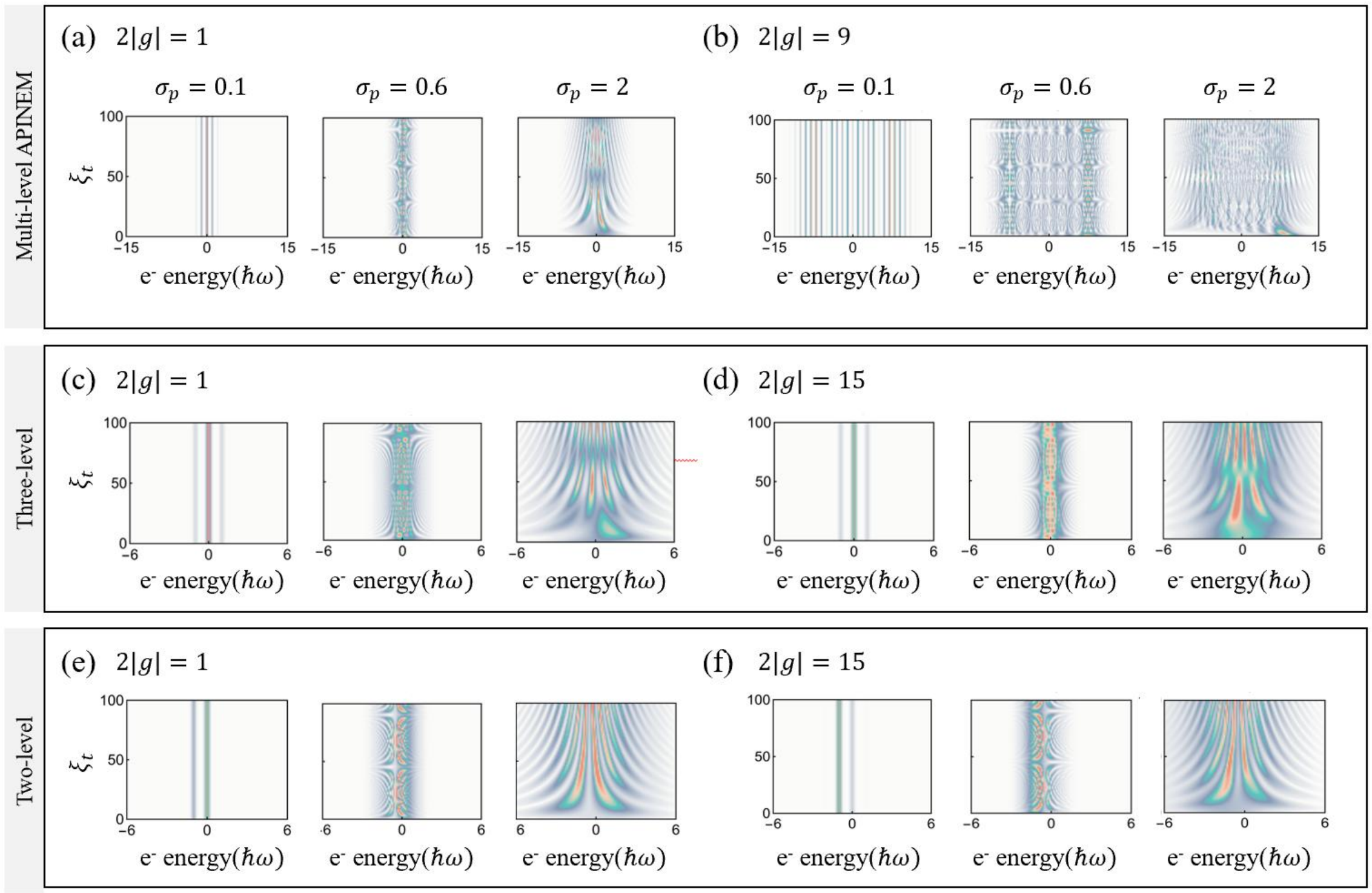

**Figure 5** | Pre-chirp control of APINEM spectra across different LEM regimes. The maps show APINEM spectra as functions of the initial momentum spread $\sigma_p$, chirping length $\xi_t$, and coupling strength $g$, with the electron energy shift normalized by $\hbar\omega$. In swift electrons, APINEM is embedded in a multilevel PINEM-like ladder, whereas in slow electrons, recoil and phase detuning confine the dynamics to three-level or two-level Bragg-regime subspaces. Increasing $\sigma_p$ and $\xi_t$ drives the spectra from narrow sideband features to broadened, oscillatory, and branched interference patterns, demonstrating pre-chirping as a control knob for APINEM engineering across Raman–Nath and Bragg regimes.

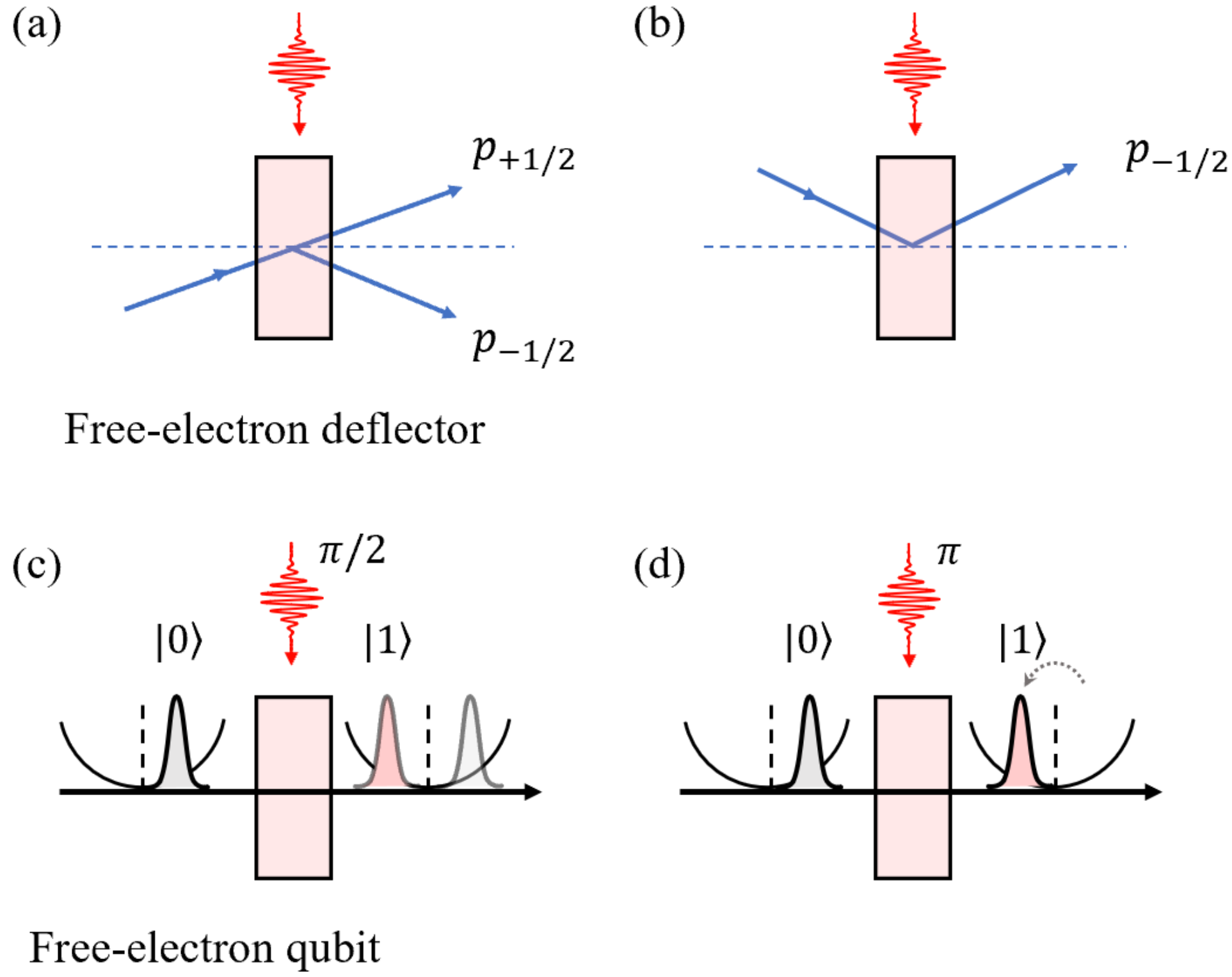

**Figure 6 |** Free-electron quantum deflector and qubit in the Bragg diffraction regime. (a,b) Bragg-matched optical coupling coherently connects two selected momentum channels, $p_{+1/2}$ and $p_{-1/2}$, realizing a controllable free-electron quantum deflector. Reversing the incident geometry reverses the momentum transfer and switches the output pathway. (c,d) The same two-state Bragg subspace encodes a free-electron qubit, with $|0\rangle$ and $|1\rangle$ denoting the two momentum states. A $\pi/2$ pulse prepares a coherent superposition, while a $\pi$ pulse inverts the qubit state. Black curves show the momentum-state wave packets, red shading marks the optically addressed components, and gray profiles/arrows indicate coherent state transformation.

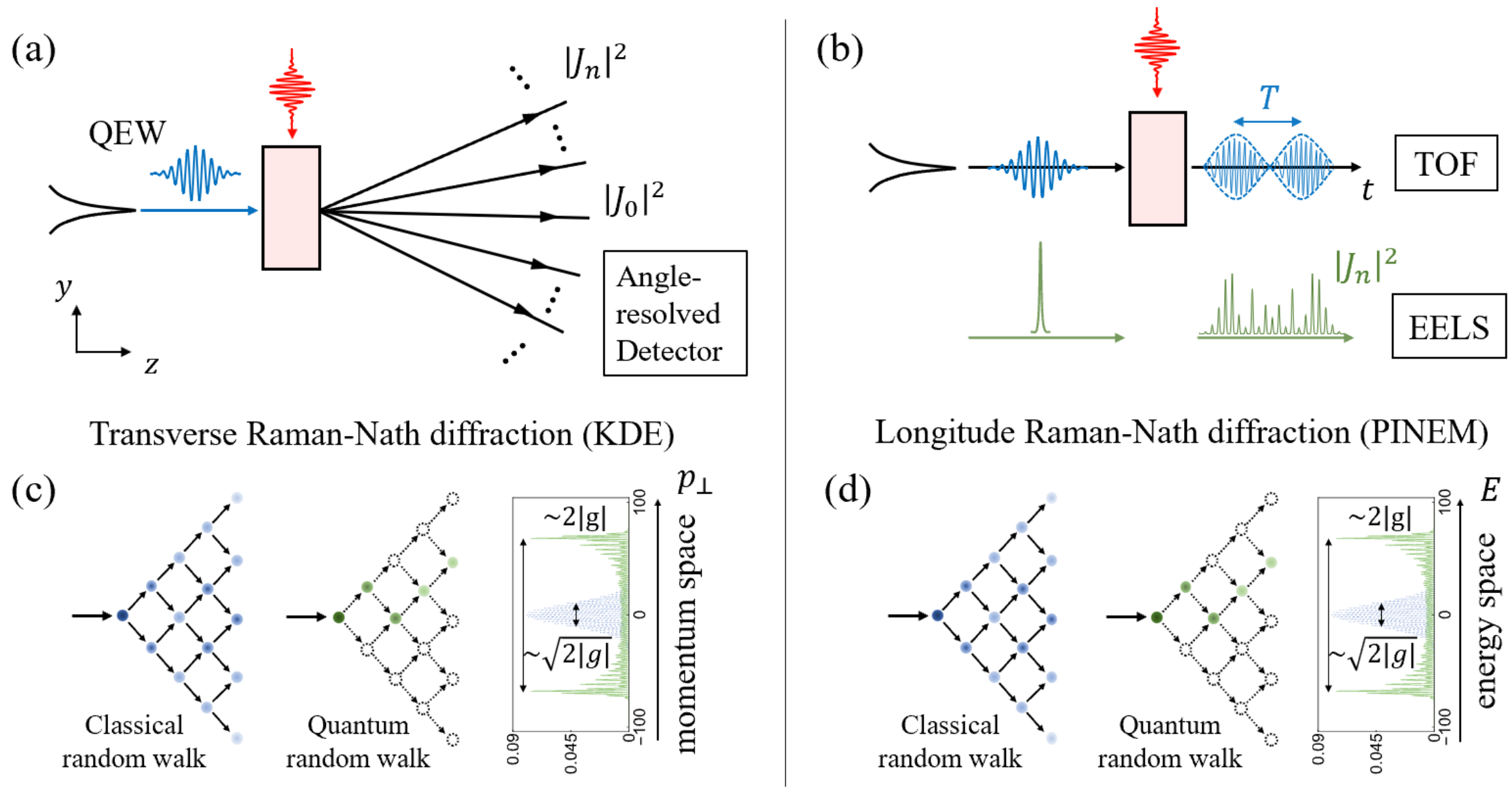

**Figure 7.** Transverse versus longitudinal Raman–Nath modulation. (a) In Kapitza–Dirac-effect architectures, transverse ponderomotive coupling scatters the electron into discrete angular diffraction orders with Bessel-governed probabilities $J_n{}^2$. (b) In PINEM configurations, longitudinal coupling generates a temporally modulated wavepacket and discrete energy sidebands spaced by $\hbar\Omega$, which can be measured by time-of-flight and electron energy-loss spectroscopy. Across both geometries, coherent QEWs undergo deterministic quantum random walks in transverse momentum or longitudinal energy space, whereas incoherent classical ensembles exhibit classical random-walk diffusion.

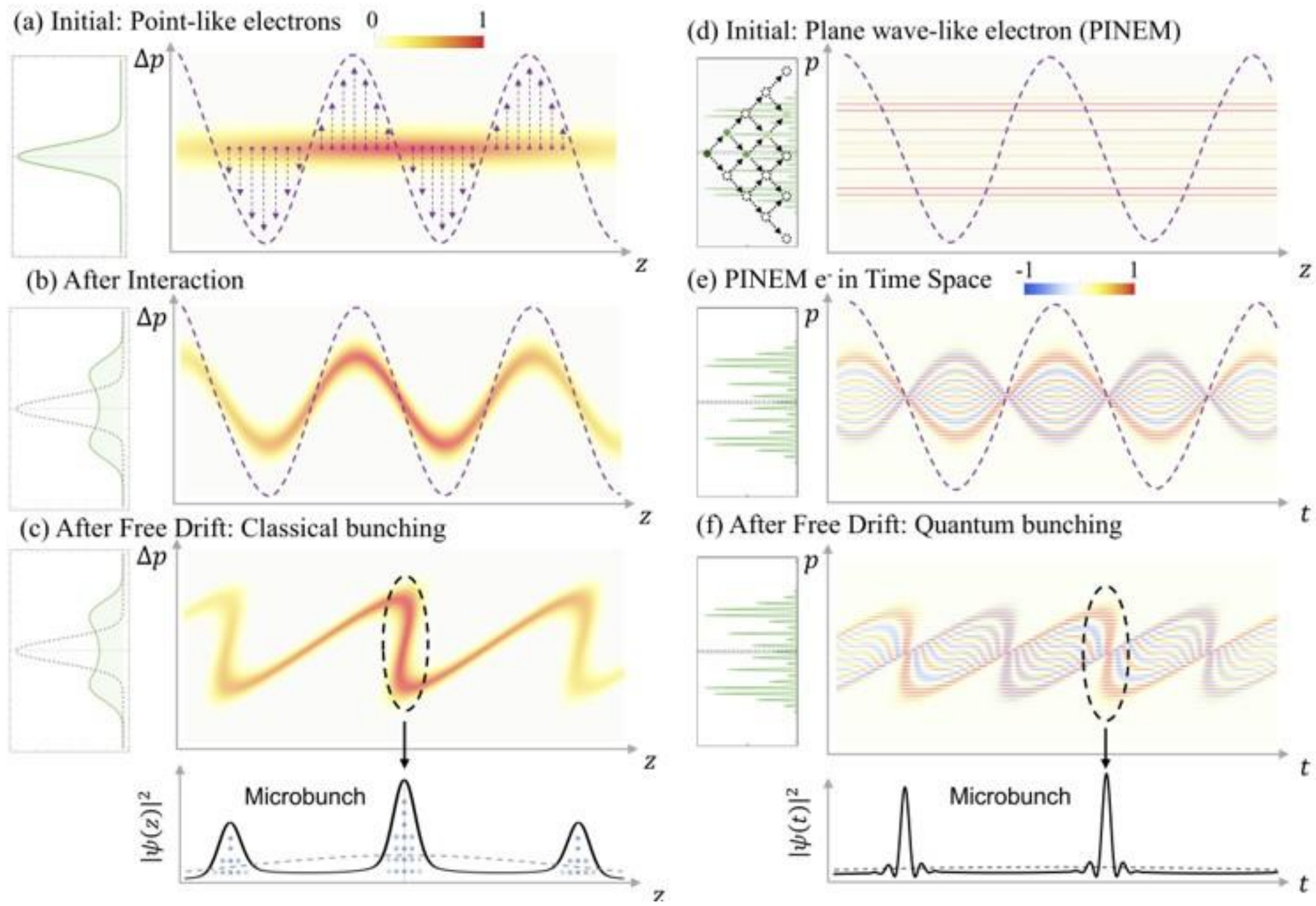

**Figure 8.** Classical density bunching versus attosecond quantum density bunching. (a–c) In the classical DLA limit, the optical near field imprints a continuous sinusoidal velocity chirp onto an ensemble of localized electrons. During subsequent free drift, this chirp drives phase-space shear and ballistic trajectory convergence, producing classical microbunches. (d–f) In the extended plane-wave PINEM limit, optical modulation coherently populates a discrete multiphoton sideband comb of a single-electron wavepacket. Free-space propagation accumulates differential phase between these sidebands, producing temporal beating and synthesizing attosecond quantum microbunches. Although both limits generate periodic density trains, their mechanisms are fundamentally distinct: classical particle-trajectory focusing in the DLA limit versus coherent multiphoton matter-wave interference in the PINEM limit.

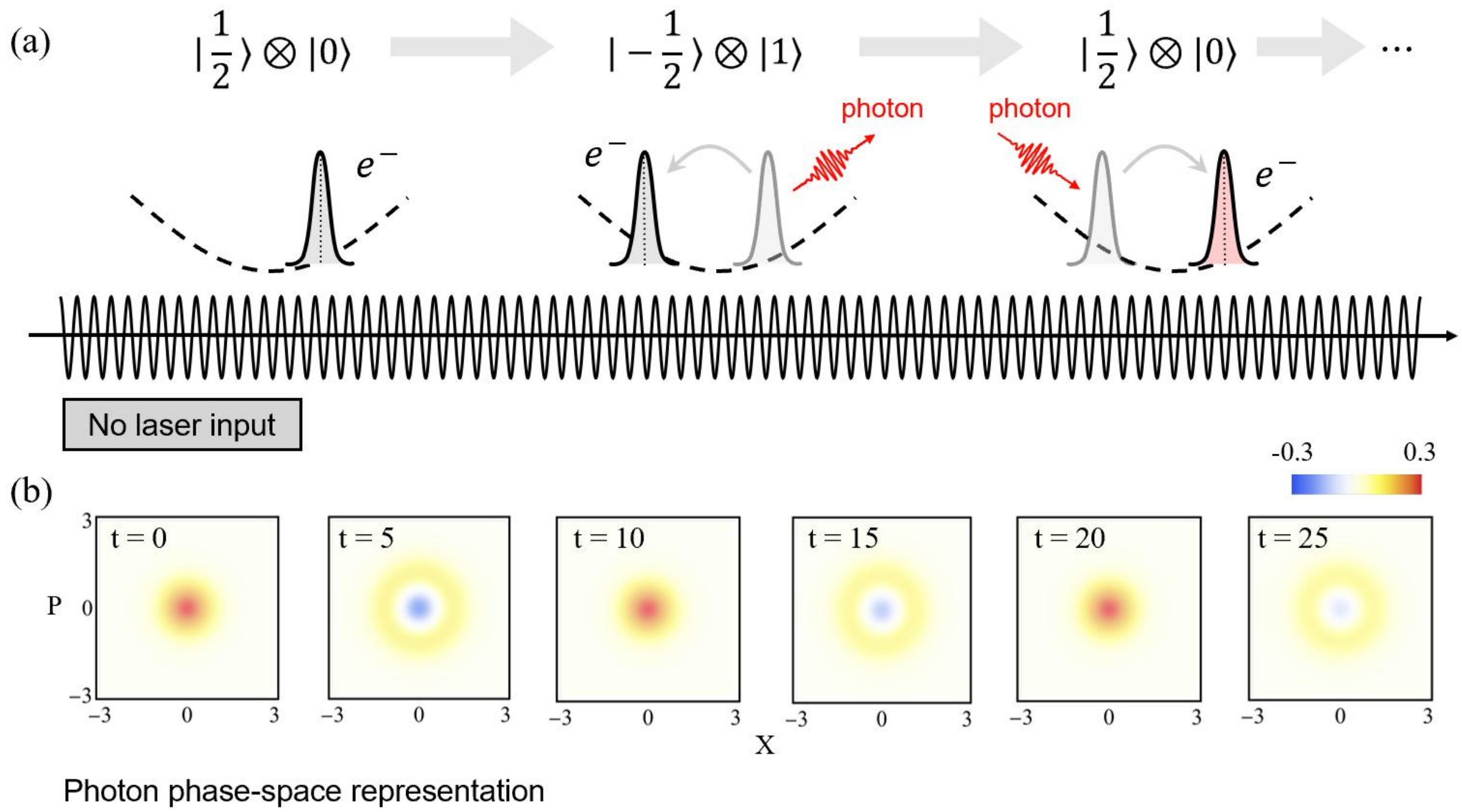

**Figure 9.** Free-electron vacuum Rabi oscillation driven by spontaneous radiation reaction. The figure shows a laser-free, cavity-mediated coherent exchange between a free-electron two-level subspace and a single quantized photonic mode. The upper row depicts reversible state evolution between the electron state and the photon-number state. The middle schematic illustrates the physical mechanism: without an external laser input, the electron spontaneously emits into and reabsorbs from the cavity mode through radiation reaction, generating a vacuum-Rabi-type oscillation between the dressed electron states. The lower row shows the photon phase-space distribution at representative times t = 0, 5, 10, 15, 20, and 25, revealing periodic buildup, depletion, and revival of the photonic excitation during coherent electron–photon energy exchange.

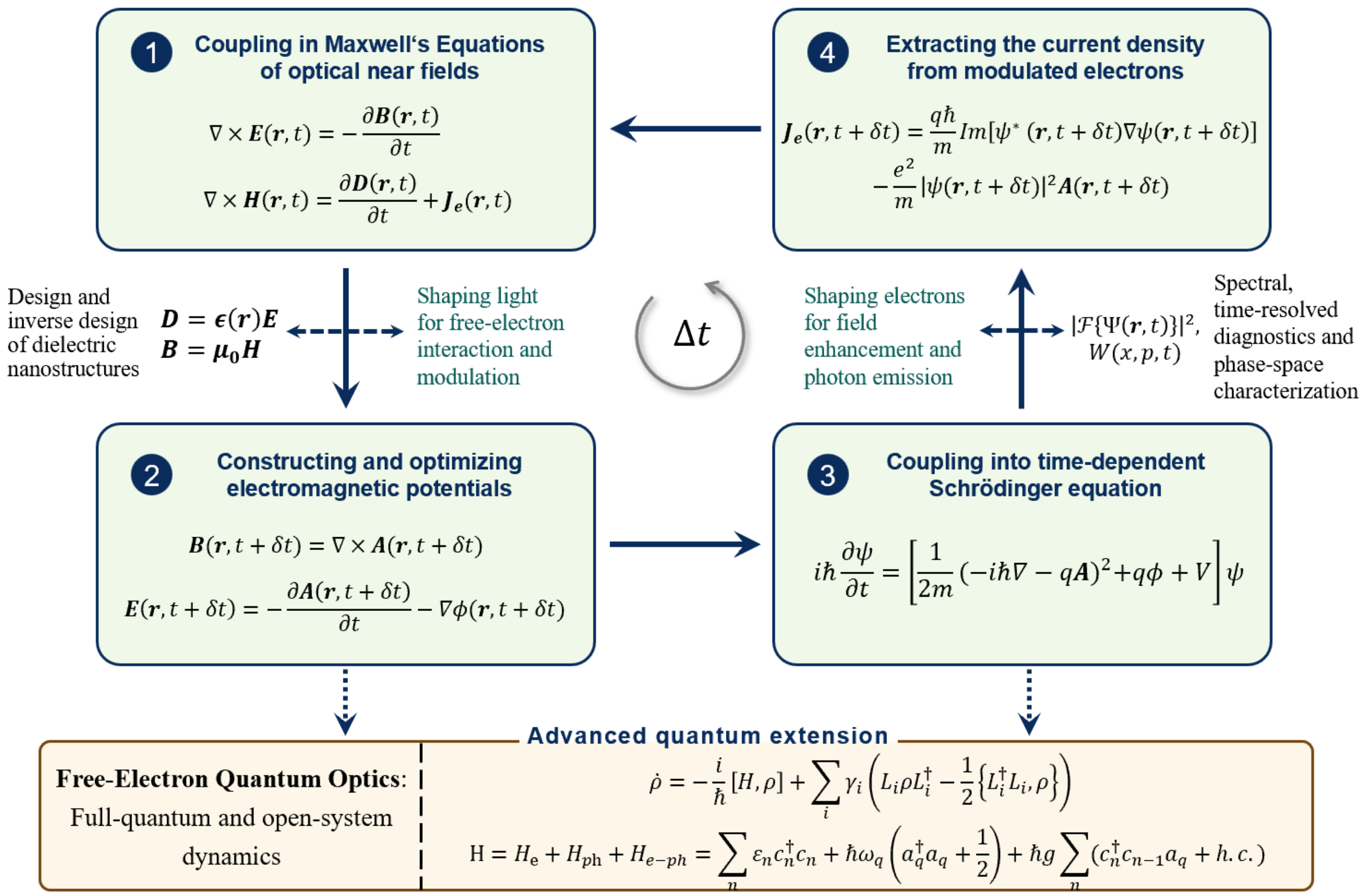

**Figure 10.** Self-consistent Maxwell–Schrödinger workflow for micro/nanophotonic LEM design. The upper panel presents a closed-loop semiclassical simulation framework. A nanophotonic structure defines the optical near field and the electromagnetic potentials ($\boldsymbol{A}, \phi$), which drive the time-dependent Schrödinger evolution of the quantum electron wavepacket. The evolved wavepacket produces a probability current density $\boldsymbol{J}_e$ that is fed back into Maxwell's equations as a reciprocal source term, enabling self-consistent optimization of field confinement, phase-matched coupling, multiphoton wavepacket shaping, and coherent photon emission. The workflow yields temporal, spectral, and phase-space diagnostics for evaluating LEM performance. The lower panel extends this framework to open-system free-electron quantum optics by replacing semiclassical fields with a quantized bipartite electron–photon Hamiltonian and evolving the density matrix with a Lindblad master equation that includes cavity loss, optical dephasing, and matter-wave decoherence.

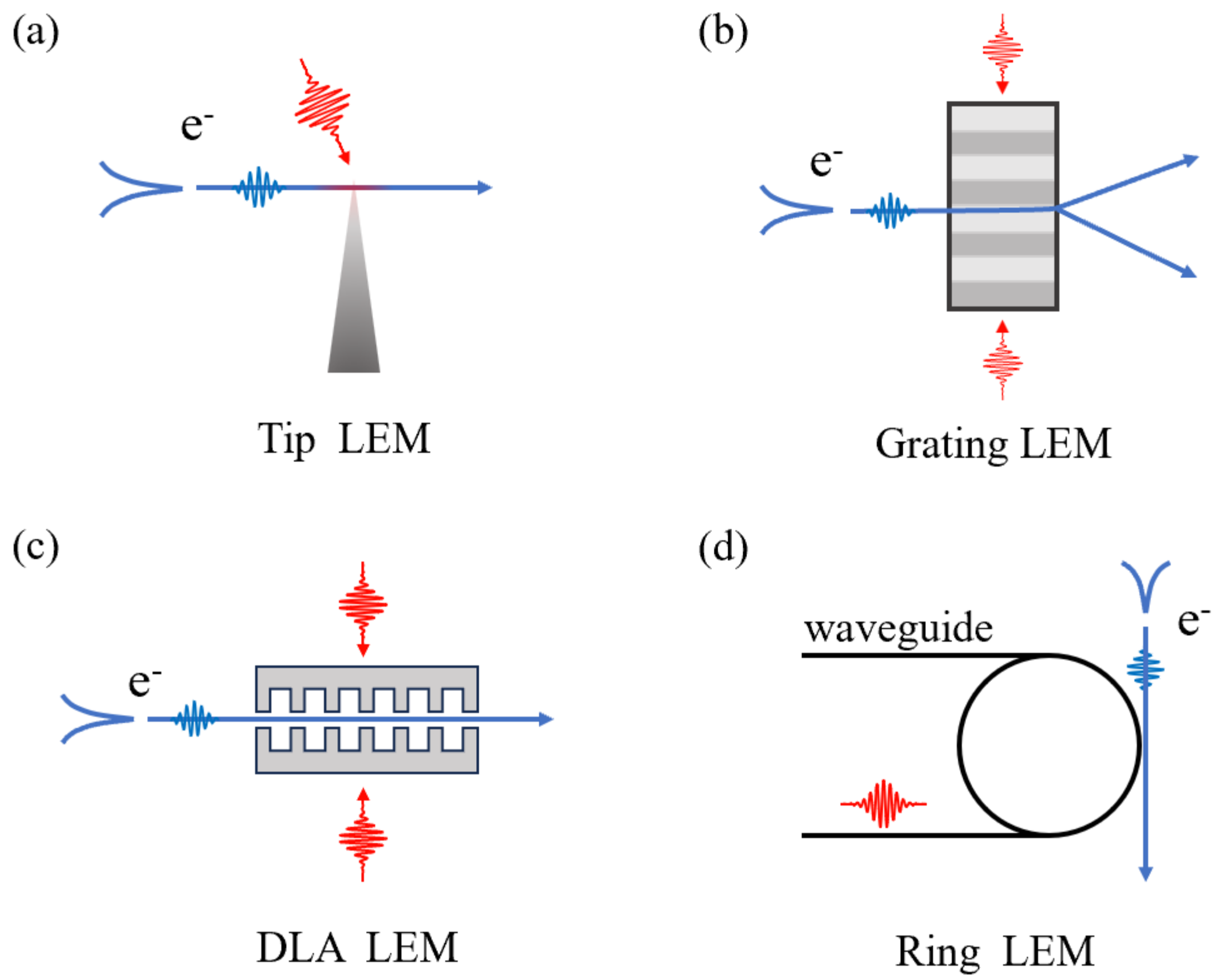

**Figure 11.** Representative device architectures for light–electron modulators. (a) Tip LEM: a plasmonic hotspot at the emitter apex provides subwavelength near-field confinement for photon-assisted wavepacket shaping. (b) Grating LEM: a laser-illuminated periodic lattice generates an extended near field that diffracts the electron into discrete transverse momentum channels. (c) DLA LEM: a dielectric vacuum channel supports phase-synchronous longitudinal coupling for kinetic-energy modulation, acceleration, and temporal bunching. (d) Ring LEM: a high-Q guided or resonant optical mode provides evanescent coupling in an integrated photonic geometry, enabling efficient continuous-wave modulation through an extended interaction length.